 \documentclass[journal,compsoc,final,a4paper]{IEEEtran}
\ifCLASSINFOpdf
\else
\fi
\usepackage{fancyhdr, epsfig, epsf, amsthm, amsmath, amssymb, amsfonts, dsfont, subfigure,color,xcolor}
\usepackage{soul}
\usepackage{enumerate,enumitem}
\usepackage{cite}
\usepackage{comment}
\usepackage{bm}
\usepackage{balance}
 \usepackage[english]{babel}

\usepackage{mathtools}
\usepackage{graphicx}
\usepackage{bbm}
\newtheorem{theorem}{Theorem}
\usepackage{cases}
\usepackage{booktabs,chemformula}

\usepackage[shortcuts,acronym]{glossaries}

\newcounter{uc}
\newcommand{\usecase}[1]{\stepcounter{uc}\noindent \textbf{Use case} \theuc. \textbf{\textit{#1}}}
\newcounter{enablerRel}
\newcounter{enablerLat}
\newcommand{\enablerRel}[1]{\stepcounter{enablerRel}\noindent \textbf{High Reliability Enabler} \theenablerRel. \textbf{\textit{#1}}}
\newcommand{\enablerLat}[1]{\stepcounter{enablerLat}\noindent \textbf{Low Latency Enabler} \theenablerLat. \textbf{\textit{#1}}}
\glsdisablehyper 

\newcommand{\CEN}{\texttt{CEN}}
\newcommand{\ASYNC}{\texttt{extFL}}

\newcommand{\queue}{q}
\newcommand{\queueTH}{\queue_0}
\newcommand{\Proposed}{\texttt{PROPOSED}}
\newcommand{\FirstBaseline}{\texttt{BASELINE-1}}
\newcommand{\SecondBaseline}{\texttt{BASELINE-2}}

\newacronym{2d}{2D}{2-dimensional}
\newacronym{3d}{3D}{3-dimensional}
\newacronym{3dof}{3DoF}{3 degrees-of-freedom}
\newacronym{3gpp}{3GPP}{Third Partnership Project}
\newacronym{4g}{4G}{fourth generation}
\newacronym{5g}{5G}{fifth generation}
\newacronym{5gnr}{5G NR}{5G New Radio}
\newacronym{5gnsa}{5G NSA}{5G non-standalone}
\newacronym{5gsa}{5G SA}{5G standalone}
\newacronym{5gppp}{5GPPP}{5G Public Private Partnership}
\newacronym{6dof}{6DoF}{6~degrees-of-freedom}
\newacronym{adas}{ADAS}{advanced driver-assistance system}
\newacronym{ai}{AI}{artificial intelligence}
\newacronym{ann}{ANN}{artificial neural network}
\newacronym{aoi}{AoI}{age of information}
\newacronym{ar}{AR}{augmented reality}
\newacronym{arvr}{AR/VR}{virtual and augmented reality}
\newacronym{asic}{ASIC}{application-specific integrated circuit}
\newacronym{bf}{BF}{beamforming}
\newacronym{bptt}{BPTT}{backpropagation through time}
\newacronym{bs}{BS}{base station}
\newacronym{c3}{C$^3$}{Communications, Computing and Caching}
\newacronym{ca}{CA}{carrier aggregation }
\newacronym{cdf}{CDF}{cumulative density function}
\newacronym{ci}{CI}{computational intelligence}
\newacronym{cn}{CN}{core network}
\newacronym{comp}{CoMP}{coordinated multi-point}
\newacronym{cpu}{CPU}{central processing unit}
\newacronym{csi}{CSI}{channel state information}
\newacronym{d2d}{D2D}{device-to-device}
\newacronym{da}{DA}{deferred acceptance}
\newacronym{dc}{DC}{dual connectivity}
\newacronym{dl}{DL}{downlink}
\newacronym{dnn}{DNN}{deep neural network}
\newacronym{dqn}{DQN}{deep-\emph{Q} network}
\newacronym{dof}{DoF}{degrees of freedom}
\newacronym{dpp}{DPP}{drift-plus-penalty}
\newacronym{drnn}{DRNN}{deep recurrent neural network}
\newacronym{dsrc}{DSRC}{dynamic short range communication}
\newacronym{e2e}{E2E}{end-to-end}
\newacronym{ea}{EA}{evolutionary algorithm}
\newacronym{ec}{EC}{evolutionary computation}
\newacronym{ehf}{EHF}{extremely high frequency}
\newacronym{embb}{eMBB}{enhanced mobile broadband}
\newacronym{eqr}{EQR}{equirectangular}
\newacronym{esi}{ESI}{extended sensed information}
\newacronym{etsi}{ETSI}{European Telecommunications Standards Institute}
\newacronym{evt}{EVT}{extreme value theory}
\newacronym[longplural=evolution strategies]{es}{ES}{evolution strategy}
\newacronym{fl}{FL}{federated learning}
\newacronym{fov}{FoV}{field of view}
\newacronym{ga}{GA}{genetic algorithm}
\newacronym{gdr}{GDR}{global discrete recombination}
\newacronym{gpu}{GPU}{graphical processing unit}
\newacronym{gru}{GRU}{gated recurrent unit}
\newacronym{hd}{HD}{high definition}
\newacronym{hdmi}{HDMI}{High-Definition Multimedia Interface}
\newacronym{hetnet}{HetNet}{heterogeneous network}
\newacronym{hevc}{HEVC}{high efficiency video coding}
\newacronym[longplural=harmony memories]{hm}{HM}{harmony memory}
\newacronym{hmcr}{HMCR}{harmony memory considering rate}
\newacronym{hmd}{HMD}{head mounted device}
\newacronym{hrllbb}{HRLLBB}{highly reliable low latency broadband}
\newacronym{hsa}{HSA}{harmony search algorithm}
\newacronym{ia}{IA}{initial access}
\newacronym{ici}{ICI}{inter-cell interference}
\newacronym{i.i.d.}{i.i.d.}{independent and identically distributed}
\newacronym{iiot}{IIoT}{industrial Internet of Things}
\newacronym{if}{IF}{impact factor}
\newacronym{imt}{IMT}{International Mobile Communications}
\newacronym{iov}{IoV}{Internet of Vehicles}
\newacronym{iot}{IoT}{Internet of Things}
\newacronym{isi}{ISI}{inter-symbol interference}
\newacronym{its}{ITS}{intelligent transportation system}
\newacronym{itu}{ITU}{International Telecommunications Union}
\newacronym{jcr}{JCR}{Journal Citation Reports}
\newacronym{kpi}{KPI}{key performance indicator}
\newacronym{lidar}{LIDAR}{laser imaging detection and ranging}
\newacronym{los}{LOS}{line-of-sight}
\newacronym{lte}{LTE}{Long Term Evolution}
\newacronym{lq}{LQ}{low quality}
\newacronym{MandE}{M\&E}{Media \&  Entertainment}
\newacronym{mac}{MAC}{medium access control}
\newacronym{mcc}{MCC}{mobile cloud computing}
\newacronym{mfg}{MFG}{mean-field game}
\newacronym{mxconn}{MC}{multi-connectivity}
\newacronym{mec}{MEC}{multi-access edge computing}
\newacronym{mimo}{MIMO}{massive input massive output}
\newacronym{ml}{ML}{machine learning}
\newacronym{mmap}{mmAP}{millimeter wave access point}
\newacronym{mmhmd}{mmHMD}{millimeter wave head mounted device}
\newacronym{mle}{MLE}{maximum likelihood estimate}
\newacronym{msi}{MSI}{model state information}
\newacronym{mtc}{MTC}{machine type communications}
\newacronym{mmtc}{mMTC}{massive machine type communications}
\newacronym{mmwave}{mmWave}{millimeter wave}
\newacronym{mno}{MNO}{mobile network operator}
\newacronym{mr}{MR}{mixed reality}
\newacronym{mtp}{MTP}{motion-to-photon}
\newacronym{mue}{MUE}{mobile user equipment}
\newacronym{nfv}{NFV}{network function virtualization}
\newacronym{nlos}{NLOS}{non-line-of-sight}
\newacronym{nn}{NN}{neural network}
\newacronym{noma}{NOMA}{non-ortogonal multiple access}
\newacronym{par}{PAR}{Pitch Adjustment Rate}
\newacronym{per}{PER}{packet error rate }
\newacronym{pf}{PF}{proportional fair}
\newacronym{phy}{PHY}{physical layer}
\newacronym{pso}{PSO}{Particle Swarm Optimization}
\newacronym{qoe}{QoE}{quality-of-experience}
\newacronym{qos}{QoS}{quality-of-service}
\newacronym{qsi}{QSI}{queue state information}
\newacronym{ran}{RAN}{radio access network}
\newacronym{relu}{ReLU}{rectified linear unit}
\newacronym{rf}{RF}{radio-frequency}
\newacronym{rl}{RL}{reinforcement learning}
\newacronym{rlf}{RLF}{radio-link failure}
\newacronym{rnn}{RNN}{recurrent neural network}
\newacronym{rrm}{RRM}{radio resource management}
\newacronym{rsr}{RSR}{Random Selection Rate}
\newacronym{rss}{RSS}{received signal strength}
\newacronym{rsu}{RSU}{road side unit}
\newacronym{sa}{SA}{simulated annealing}
\newacronym{sbs}{SBS}{small cell base station}
\newacronym{scn}{SCN}{small cell network}
\newacronym{sdn}{SDN}{software defined networking}
\newacronym{sdtv}{SDTV}{standard-definition television}
\newacronym{sfn}{SFN}{single frequency network}
\newacronym{sinr}{SINR}{signal-to-interference-plus-noise ratio}
\newacronym{snr}{SNR}{signal-to-noise ratio}
\newacronym{so}{SO}{stochastic optimization}
\newacronym{soa}{SOA}{state-of-the-art}
\newacronym{son}{SON}{self-organizing network}
\newacronym{sue}{SUE}{small cell user equipment}
\newacronym{tdd}{TDD}{time division duplexing}
\newacronym{uav}{UAV}{unmanned autonomous vehicle}
\newacronym{udn}{UDN}{ultra dense network}
\newacronym{ue}{UE}{user equipment}
\newacronym{uhd}{UHD}{Ultra High Definition}
\newacronym{ul}{UL}{uplink}
\newacronym{urllc}{URLLC}{ultra reliable low latency communication}
\newacronym{v2i}{V2I}{vehicle-to-infrastructure}
\newacronym{v2v}{V2V}{vehicle-to-vehicle}
\newacronym{v2x}{V2X}{vehicle-to-everything}
\newacronym{vor}{VOR}{vestibulo-ocular reflex}
\newacronym{vr}{VR}{virtual reality}
\newacronym{vrp}{VRP}{virtual reality player}
\newacronym{vrx}{vRx}{vehicular receiver}
\newacronym{vtx}{vTx}{vehicular transmitter}
\newacronym{vue}{vUE}{vehicular user equipment}
\newacronym{wi-fi}{WI-FI}{wireless-fidelity}
\newacronym{xr}{XR}{extended reality}

\begin{document}

\title{Wireless Edge Computing with Latency \\ and Reliability Guarantees}
\author{Mohammed~S.~Elbamby,
        Cristina~Perfecto,
        Chen-Feng Liu,~\IEEEmembership{Student~Member,~IEEE,}\\
        Jihong~Park,
        Sumudu~Samarakoon,
        Xianfu Chen,~\IEEEmembership{Member,~IEEE}
        and~Mehdi~Bennis,~\IEEEmembership{Senior~Member,~IEEE}%
        
    \thanks{Mohammed~S.~Elbamby, Chen-Feng Liu, Jihong~Park, and Sumudu~Samarakoon are with the Centre for Wireless Communications, University of Oulu, 90014 Oulu, Finland (emails: firstname.surname@oulu.fi).}
     \thanks{ Mehdi~Bennis is with the Centre for Wireless Communications, University of Oulu, 90014 Oulu, Finland, and also with the Department of Computer Science and Engineering, Kyung Hee University, Seoul 17104, South Korea (e-mail: mehdi.bennis@oulu.ﬁ).}

    \thanks{Cristina~Perfecto, is with the University of the Basque Country (UPV/EHU), Spain. (email: cristina.perfecto@ehu.eus)}
    \thanks{Xianfu~Chen is with VTT Technical Research Centre of Finland, P.O. Box 1100, FI-90571 Oulu, Finland (e-mail: xianfu.chen@vtt.fi).}%
\vspace{-0.4cm}}
\maketitle
\begin{abstract}
Edge computing is an emerging concept based on distributing computing, storage, and control services closer to end network nodes. Edge computing lies at the heart of the \gls{5g} wireless systems and beyond. While current state-of-the-art networks communicate, compute, and process data in a centralized manner (at the cloud), for \emph{latency and compute-centric} applications, both radio access and computational resources must be brought closer to the edge, harnessing the availability of  computing and storage-enabled small cell base stations in proximity to the end devices. Furthermore, the network infrastructure must enable a distributed edge decision-making service that learns to adapt to the network dynamics with minimal latency and optimize network deployment and operation accordingly.  This article will provide a fresh look to the concept of edge computing by first discussing the applications that the network edge must provide, with a special emphasis on the ensuing challenges in enabling ultra-reliable and low-latency edge computing services for mission-critical applications such as \gls{vr}, \gls{v2x}, edge \gls{ai}, and so forth. Furthermore, several case studies where the edge is key are explored followed by insights and prospect for future work. 
\end{abstract}
\IEEEpeerreviewmaketitle
%
\glsresetall 
\section{Introduction}

\IEEEPARstart{T}{he} ever increasing requirements of wireless services in \gls{MandE}, as well as in healthcare and wellbeing demands are transforming the way data is communicated and processed. Future networks are anticipated to support massive number of connected devices requesting a variety of different services such as mobile video streaming, \gls{arvr}, as well as mission-critical applications. Such services require data, computation, and storage to be performed more often with ultra-high success rate and minimal latency. \Gls{mec} has emerged as an infrastructure that enables data processing and storage at the network edge as a means to cut down the latency between the network nodes and the remote servers that typically existed in cloud computing architectures \cite{ShiEC16}. Instead, edge computing can be provided as a service at the network edge to minimize the service latency, network complexity, and save the device nodes' energy and battery consumption.

 Edge networking in cellular systems aims to efficiently provide the required connectivity, data access, bandwidth, and computation resources to end devices \cite{VodafoneEdgeCellular, SaadEC18}. Edge base stations in proximity of network users will not only relay content from and to the network core, but will help execute the users processing tasks, provide customized content and computing services, and control the connectivity and interaction between coupled network nodes.

 
 In essence, the performance of edge computing is predominantly assessed through two main components, communication between the edge server and  the end device, and the processing at the edge server. Further, several optimization aspects are considered to optimize these two components. Optimizing the communication part can be explored through wireless bandwidth and power allocation, edge server selection, computation task distribution, task splitting, and partial task offloading. For the processing part, computation cycle allocation, task queuing and prioritization, joint computing, and predictive computing are critical factors to optimize the computing efficiency. 
 

The focus of the \gls{5g} cellular networks has shifted from merely increasing the data communication rate to providing service-specific performance guarantees in terms of ultra-reliability and low latency. This shift is fueled by the emergence of new use cases that require genuine support to critical and latency-sensitive communication services.  Nonetheless, ultra-reliability and low latency are often seen as contradictory requirements \cite{conf:latencyReliability_soret_2014}, compelling the use of distinctive set of tools to be efficiently realized.  Yet, these individually challenging per se requirements are anticipated to be met together for networks of diverse topologies and heterogeneous services.


This article discusses the feasibility and potential of providing edge computing services with latency and reliability guarantees. In particular, it first sheds light on the services that can be offered from edge computing networks. It follows by looking into how \gls{urllc} contributes to and benefits from edge computing. The article proceeds by presenting selected use cases that reflect the interplay between edge computing and \gls{urllc}. Finally, the article ends with our concluding remarks and future works.

\section{Edge Computing Services}
Legacy network architectures relied on centrally located and centrally-controlled servers with high computational and storage powers to provide on-demand computing to network devices \cite{Kaibin_MEC_survey}. These servers could support high number of network nodes over a large geographical area. However, the large distance between the cloud computing server and end-user device results in higher service latency. Moreover, the centralized architecture limited the ability to provide context-aware service, and to preserve the user data privacy. Future wireless networks are evolving towards supporting new set of applications that require minimal latency and high level of service personalization. This motivated the shift towards distributed networking architectures where the network resources are available close to users at the network edge. Edge computing aims to provide computing, content, and connectivity services closer to the data source and consumption points.  
It is applicable to scenarios with different network environments and use cases. This diversity led to several implementations that did not follow specific standard or interoperability. The \gls{etsi} has been working on solving this issue through providing an efficient standardized \gls{mec} that can be integrated across several applications and service providers \cite{etsi_mec}. \gls{mec} also enables providers to deploy edge computing services on top of wireless mobile networks. This will allow cellular operators to integrate computing into the services provided to their users. In this regard, the term \emph{edge networking} refers to the action and process of serving a user or device at the network edge.


\subsection{Content at the Edge}
The idea of leveraging the network edge as a content storage has gained popularity in the last few years \cite{Bastug_caching}. Existing popularity patterns on the contents requested by network users motivated developing proactive networks.  A proactive server can predict popular contents, prefetch them from the core network, and have them stored and readily available at the network edge, hence cutting down delivery times  once users request them.  Proactive networks require efficient methods to predict the popularity of the content to be cached, as well as high storage capacity to cache this content. Edge caching not only minimizes the service latency but also the load on the backhaul network by prefetching the popular content in the off-peak times \cite{Bastug14,Bennis16,Elbamby14}. 
Further, we envision that the notion of edge content will be extended to include new types of data that can be served from the network edge to support the new use cases. One application to which the future network edge will provide information is the distributed machine learning application. The tight latency requirements and the need for minimizing the information exchange mandate the development of distributed machine intelligence schemes in which edge servers play a major rule. Edge machine learning \cite{Park:2018aa, Li_edgeInt18} will allow end users to locally develop their own machine learning models instead of relying on centralized approaches. However, "machine learning applications" rely on information from other network nodes that affect their state and utility. The network edge role here will be to bring the information necessary for enhancing or complementing the local model close to the user.

\subsection{Computing at the Edge}
Processing is becoming as an important commodity to cellular applications as content. The use of applications ranging from smart factory, self-driving vehicles, to virtual and augmented reality are growing by the day and are becoming more resource greedy and less latency tolerant. While part of the computing load of these applications is served using their local processing units, constraints on size, portability, battery life-time or lack of full access to task data limit the ability to locally execute computing tasks. Edge computing promises to pool powerful yet proximate computing resources at the network edge, as well as to provide connectivity and seamless information exchange between neighboring nodes. It is also set to allow for the realization of various \gls{5g} verticals that require low-latency and high-reliable computing, such as \gls{vr} and mission-critical \gls{iot} applications. Yet, there are several components that need to be addressed to realize low-latency and high-reliable edge computing. Executing computing tasks at the edge often requires the task data to be offloaded to the edge server before execution. This introduces communication delay that adds to the service latency. In addition, how to queue and schedule the computing tasks at the edge server plays a major role in the queuing and processing latency.  Our vision is that the availability of more data and computing power will shape how the edge network performs computing. Similar in vein to proactive content caching, where knowledge of users’ preferences and future interests allow for prefeching of their content, data availability and machine learning will help to speed up computing the tasks of network nodes. Predicting vehicles’ future locations and path allows the edge network to proactively render and deliver its high-definition live map. In \gls{vr} applications, predicting users’ future \gls{fov} allows rendering the corresponding part of its $360^{\circ}$ frame with minimal latency. Several other enablers are vital to achieve ultra-reliable and low-latency computing, such as task replications, parallel, and coded computing, which will be addressed in detail in the following section.    
 \begin{figure*}[ht]
    \centering
	\includegraphics[width=.92\textwidth]{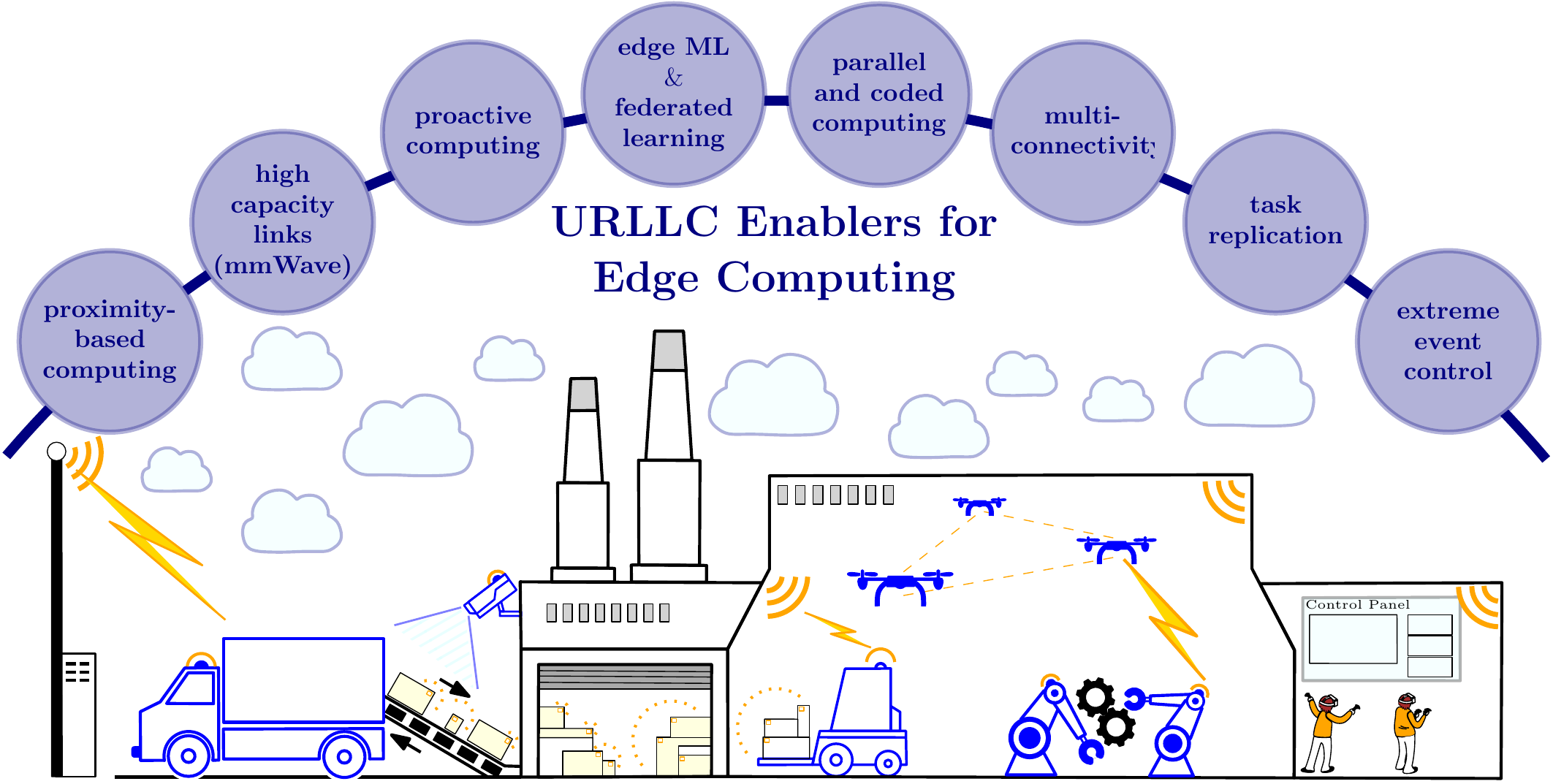}
	\caption{Breakdown of key URLLC enablers for edge computing, exemplified over an Industry 4.0/Smart Factory ecosystem that includes cyber-physical systems, \gls{iot} and \gls{mec}.}
	\label{Fig:mec_env}
 \end{figure*}
\subsection{Control at the Edge}

Most of the existing cloud and edge computing architectures rely on centralized decision-making schemes which requires all the network nodes to send their local states data to a central controller. Instead, distributed decision making, in which the decision-making process is distributed among the edge servers will allow for low latency, and privacy preserving operation \cite{EdgeMesh17}, which is essential for mission-critical applications. Indeed, the control of the network devices’ performance requires policies that adapt to their local states. This can be challenging for scenarios where the local state dynamically varies due to highly dynamic environment or due to the nature of the application, such as in mission-critical applications. \Gls{rl} solutions can provide efficient control policies that maximize the system rewards by finding policies that map those dynamically changing states into actions. These decision-making policies need to take into account the effect of actions on the environment and update the reward accordingly.  In centralized architectures, classical reinforcement learning is often performed offline, not taking into account reliability in decision making for example under noisy feedback. Edge control can provide robust decision-making, where multi-agent \gls{rl} architectures can be used to provide communications efficient methods that take latency and reliability into account in dynamic and mission-critical environments. Latency stems from the local state exchanges between edge devices, in which the overhead due to the state exchange increases exponentially with the number of devices. This can be addressed using the \gls{mfg} theory~\cite{MFGJ}, which can tackle this by approximating the average state as a collection of agents' instantaneous states.    
 
\section{URLLC Enablers and Challenges}

\subsection{URLLC overview}
The prime focus of the recent groundswell of mission critical applications such as autonomous vehicles, immersive \gls{vr}/\glsunset{ar}\gls{ar} experiences, industrial automation, and robotics, is to provide services with guaranteed high reliability and low latency.
Therein, latency deductions in channel estimations, information exchange among the network elements, decision making, computation tasks completion, and memory access within devices have utmost importance.
Along with them, guaranteed low-latency in operations, ensuring connectivity, and speed-precision-and-accuracy of computations are essential to assure the reliability of mission critical applications. 
Due to the on-device constraints on storage, processing capability, and availability and accessibility of network resources, it is mandatory to utilize the edge servers to maintain the quality-of-service in mission critical applications.
To support the communication among user devices within mission critical applications and the edge servers, \gls{urllc}, that has been introduced as one of the main service in 5G systems, plays a pivotal role. In this section, we identify the key enablers of reliability and low-latency in wireless edge computing networks, and the challenges towards realizing each of them. Moreover, in Table \ref{tab:urllc_mec}, we summarize the issues and enablers of providing latency and reliability guarantees in wireless edge computing networks, as well as the applications and use cases these enablers are targeting. 

\begin{table*}[]
	\centering
	\resizebox{\textwidth}{!}{\begin{tabular}{p{1cm}p{3.3cm}p{4.4cm}p{4.9cm}}
		&   Demands/Challenges & Enablers  & MEC applications and use cases  \\[3pt]
		\hline
		\textbf{Low latency} & bandwidth, backhauls & mmWave        & extended reality, vehicular edge computing (Sec. \ref{usecaseVR} and Sec.~\ref{usecaseV2V})\\[3pt]
		&  low propagation delay &  proximity based computing  &    deep reinforcement learning based task offloading (Sec.~\ref{usecaseDQN}, use case 4)\\[3pt]
		& computing power, task dependency     &    parallel and coded computing &   \cite{Chen_EdgeIoT18,CodedMR15} \\[12pt]
		& low propagation delay,  energy efficiency     &        proactive computing     & use case 6 in Sec.~\ref{usecaseVR1}\\[12pt]
		& low prediction delay       &           edge machine learning       &       edge computing for federated learning (use case 1 in Sec. \ref{usecaseDistML})\\[3pt]
		\hline
		\textbf{High reliability} & channel intermittency         &               multi-connectivity, task replication     &      use case 6 in Sec.~\ref{usecaseVR1} and \cite{ElbambyEurasip2018,PerfectoVRStreaming}\\[3pt]
		& low communication cost, data privacy    &        federated learning     &edge computing for federated learning (use case 4 in Sec.~\ref{usecaseDQN})\\[3pt]
		& rare event detection      &            extreme event control     &     extreme value theoretic edge computing and vehicular federated learning (use cases 2 and 3 in Sec.~\ref{usecaseEVT})
	\end{tabular}}
 	\caption{Challenges and enablers of realizing low latency and high reliability in wireless edge computing.}
	\label{tab:urllc_mec}
\end{table*}

\subsection{URLLC Enablers for Edge Computing}

\subsubsection{Low latency Enablers}
There are several components that contribute to latency in edge networking. In this regard, enabling low latency requires several techniques to be implemented and integrated together at different levels of edge networking systems. At the communication level, proximity-based computing and \gls{mmwave} links play major roles in reducing task offloading latency from edge devices to servers by reducing distance attenuation and providing broad bandwidth with high directionality, respectively. In addition, \gls{mmwave} also enables wireless backhauling \cite{jnl:mmWBackhaul_CommunMag_2014,conf:vu_mmWBackhaul_EUWireless_2016} that facilitates edge servers' prefetching popular content with low latency. At the processing level, proactive computing provides significant latency reduction while maximizing resource efficiency by avoiding repetitive and redundant on-demand computing \cite{ElbambyProactive17,Oueis16, ElbambyEurasip2018}. Next, coded computing is effective in reducing parallel computing latency, which eliminates the dependency of processing tasks, thereby minimizing the worst-case latency due to a straggling task. Last but not least, \gls{ml} is crucial in supporting low-latency mission critical applications, by empowering edge servers and devices to locally carry out their decision-making.

\vspace*{.2cm}
\enablerLat {High capacity mmWave links}:
Driven by the spectrum shortage below 6 GHz, communications in the radio frequencies encompassing the electromagnetic spectrum from 30 to 300 GHz, i.e. the \gls{mmwave} or \gls{itu}'s \gls{ehf} band, have been attracting a growing attention~\cite{jnl:mmWitWillWork_Rappaport_2013,jnl:5GWhatWillB_Andrews_2014,jnl:5disruptive5G_Boccardi_2014}, to the point of being currently considered the most important technology to achieve the 10 Gbps peak data rates foreseen for the upcoming \gls{5g} systems~\cite{jnl:mmW_FutureMobile_JSAC_2017}.
Having abundant available spectrum, the main appeal of \gls{mmwave} communications comes from the use of generous bandwidths that \textendash ranging from the 0.85GHz in the 28GHz band to 5 GHz in the 73GHz band\textendash are more than ten times greater than \gls{lte}'s 20 MHz cellular channel~\cite{jnl:mmW_5Goverview_Rappaport_2017}, and grant an important channel capacity increase~\cite{mmWaveChannelandCellularCapacity_2014}.

However, signal propagation at these frequencies is harsh and inherently different from that at the microwave band~\cite{jnl:mmWave_Survey_2018} experiencing 1) higher pathloss for equal antenna gains due to a stronger atmospheric attenuation whereby signals are more prone to being absorbed by foliage and rain, 2) higher penetration losses as \glspl{mmwave} are blocked when trying to pass through walls, buildings, or obstacles, and 3) higher transmit power consumptions than in lower bands to preserve an equal \gls{snr} unless directional antennas together with advanced signal processing that includes \gls{mimo}~\cite{jnl:mmW_MassiveMIMO_2014} and \gls{bf} techniques are used. 
Notably, due to the shorter wavelengths in \gls{mmwave} bands it is possible to pack more antennas at the transmitter and receiver devices and, thanks to the spatial degrees of freedom afforded, use analog or hybrid \gls{bf} \textendash fully digital \gls{bf} implies having one dedicated \gls{rf} chain per antenna which currently discourages its use in \glspl{mmwave} due to the unaffordable power consumption and costs\textendash~to build a radiation pattern with narrow beams which will be subsequently steered towards the receivers while the energy radiated through the sidelobes is minimized or negligible. 

To administer high capacity links with \glspl{mmwave}, transmitters' and receivers' mainlobes need to be precisely aligned towards each other if favored with a clear, unobstructed, \gls{los} path. In practice,  
when a \gls{mue} is in the connected state, \gls{ul} control channels are used to periodically feed back to the \gls{bs} its best transmit beam index; similarly \gls{dl} control channels are used to report \glspl{mue}' best transmit beams. Data transmission is then performed through the best beam pair. However, during initial access and handover, i.e. in random access, such information on the best beams is not available which hinders taking full benefit from \gls{bf}. 
Henceforth, in analog \gls{bf}, to discover and then maintain the best transmit-receive beam pairs, a series of techniques referred to as beamtraining or beamsearching, are applied. 
Then, beam tracking is performed to adapt the beamforming, e.g., due to \glspl{mue}' movement leading to transmitter-receiver beam misalignments. Nevertheless, a full new directional channel discovery process will need to be triggered if the \gls{sinr} drops below a certain threshold due to e.g., blockages and/or interference~\cite{Mattia:18}. 
As analog \gls{bf} employs a single \gls{rf} chain, it is challenging to adjust the beam to channel conditions, leading to some performance loss. Moreover, analog \gls{bf} does not provide multiplexing gains as it can only operate a single data stream. Therefore, to bring all the benefits of \gls{mmwave} while benefiting from multiplexing gains for \gls{mec}, \gls{mimo} hybrid \gls{bf} architectures, which strike a balance between performance, complexity, and power consumption, should be considered. Finally, as adaptive beamforming requires precise \gls{csi}, one of the key challenges for \gls{mmwave} to work as a low-latency enabler for \gls{mec} lies on the availability of expedited \gls{csi} acquisition schemes together with directionality-aware mobility and beam management procedures~\cite{mmWBeamManag_Tutorial_Giordani_2019}. 

In the next subsection a series of reliability enablers will be discussed to reduce the delay incurred to counteract the intermittent blockages and temporal disruptions of the \gls{mmwave} channel.
Largely, these techniques are in line with the idea of overbooking radio resources as a protection against channel vulnerability~\cite{conf:barbarossa_overbookResources_2017} or to consider risk-sensitive approaches~\cite{Vu:2018:RSRL}.

\vspace*{.2cm}
\enablerLat {Proximity-based Computing}:
Reducing the distance between the application and the \gls{mec} server is a key latency enabler.
This idea is motivated by the concept of bringing the transmitter and the receiver closer to one another yielding capacity improvements \cite{jnl:5GWhatWillB_Andrews_2014}.
With the low proximity between the application and \gls{mec} server, over-the-air latency that has a significant contribution to the \gls{e2e}, sometimes dominating over the computing latency, can be greatly minimized.
Network densification, the concept of dense deployment of small cells, remote radio units, and relay heads that has been an attractive research interest during recent years~\cite{UDN_survey,JHParkTWC:15,MobilMFGSG:GC16,MFGSG:GC16,kim:2017:MFCA,Samarakoon16}, plays a major role in proximity-based computing.
While boosting the capacity and coverage, the dense deployment of access points offers the opportunity of introducing additional computing resources at the network edge.
Henceforth, the user devices in the network are capable of uploading their computational tasks to access points and download the corresponding outputs after the processing with high data rates yielding lower latencies.
Another proximity-based computing technique is mobility assisted \gls{mec}.
Therein, networks of connected vehicles, \gls{uav}, and robots with high processing power can assist the computational tasks of the users \cite{Li:2019,Hagenauer:2017}.
The high processing power of above devices that are dedicated to users provides low computational latencies.
Moreover, their flexible connectivity with the users due to the mobility and high data rates therein due to the proximity offer lower communication latencies, yielding reduced \gls{e2e} latencies. 
Computing location swapping is another proximity-based computing method.
Therein, groups of users coexist in either physical (located close by) or virtual spaces (interact and/or share computing tasks).
In his regard, proximity alone provides low communication latency, yet could yield poorly utilized computational resources.
Combining the user groups in virtual space and their physical locations, some users can swap their associated \gls{mec} servers to improve both computing and communication latencies, resulting better \gls{e2e} performance \cite{Park:2018:VR}.
Although the proximity-based computing enables low latency in \gls{mec}, the concept itself brings up new challenges to the network design and resource optimization therein.
The increased interference is one of the challenges in both network densification and computing location swapping.
Due to the limited availability of both communication and computation resource, increased interference may degrade both uplink and downlink communication yielding increased \gls{e2e} latency~\cite{romanous15}.
In this regard, interference avoidance, management, and mitigation techniques as well as use of higher frequency channels are viable remedies.
Another challenge is the frequent handover due to the dynamics of environment and user mobility ~\cite{romanous15,arshad16}. 
%
%
While handover may incur undesirable latencies, the concept of \gls{mxconn} can be utilized, in which users receive computing assistance from several \gls{mec} servers.
%
%
%
%

\vspace*{.2cm}
\enablerLat {Edge Machine Learning}:
Inference (or prediction) capabilities with low latency is one of the main reason for \gls{ml} to be popular in \gls{mec} as well as several other communication applications such as coding, beamforming, resource optimization, caching, scheduling, routing, and security \cite{wang2018RL,PerfectoVRStreaming,kato17,mao18}.
While the majority of the ML-based communication system design literature is rooted on the centralized and offline ML techniques,
the upturn of mission critical applications for massive number of connected devices demands for the intelligence at the network edge~\cite{Park:2018aa,KaibinIntEdge18}. 
In contrast to conventional centralized \gls{ml} designs, the edge \gls{ml} is capable of generating inference within an instance at the edge devices, presenting the opportunity to greatly reduce the \gls{e2e} latency in \gls{mec} applications.
Such intelligence at the edge devices can 
1) predict the uncertainties in channel dynamics, communication and computation resource availability, interference, and network congestion at the local devices;
2) explore and learn about the network environment with minimal additional signaling overheads;
and 
3) characterize and model the network behavior in which the system performance is analyzed.
%
%
At the \gls{mec} servers, such prior knowledge provides the opportunities to smartly schedule their computing resources and share the results with the corresponding user devices. 
%
%
%
Furthermore, at the events of connectivity losses, edge \gls{ml} at the user devices allows the decision making within the devices using the forecast on system behaviors, allowing uninterrupted end-user service experiences.
This ability to operate offline/off-grid can reduce the number of latency-critical parallel tasks at the \gls{mec} server, in which network-wide end user experience is improved.
%

%
The challenge of enabling low latency in \gls{mec} via edge \gls{ml} relies on the training latency and inference accuracy therein.
In the distributed setting, each edge device lacks the access to the large global training data set, in which training over local data can degrade the inference accuracy.
To improve the inference accuracy, edge \gls{ml} devices may need often cooperation among one another or with a centralized helper, which incurs additional overheads and thus, increased training latency.
In this regard, further investigations need to be carried out to optimize the tradeoff between training latency and inference accuracy depending on the design architectures, communication models, and application requirement.

\vspace*{.2cm}
\enablerLat {Proactive Computing}:
	Although edge computing is capable of minimizing the latency induced due to the high propagation delay of cloud computing, it still experiences delay due to offloading the task data to the edge server, processing delay, as well a queuing delay for both operations. 
	While these delays are inevitable in some cases, there exists situations in which the task has already been executed before for another user at a different time.
	Take for example an \gls{ar} case in which visitors of a specific spot in an exhibition or museum request a specific task of augmenting an object to the view of this spot, or the task of object identification by multiple vehicles in an \glspl{its} system.
	Executing these tasks redundantly each time it is requested is certainly not resource efficient, and is causing higher delays to these tasks as well as other tasks sharing these resources.
	Here, executing and caching the results of these tasks in advance, such that they are served when requested with minimal latency, can be a major latency minimizer.

	The ideas of prefetching tasks \cite{KoPrefetching17} and proactive computing \cite{ElbambyProactive17,Oueis16} aim to develop techniques that learns and predicts which tasks are to be requested in the future and pre-compute them.
	Indeed, the success of proactive computing lies on a well-aimed choice of which tasks to proactively compute and which are to leave for real-time processing. Essentially, this involves developing efficient prediction methods that   studies the popularity patterns of the computing tasks to decide on which tasks to prefetch. 
	The idea also relies on the availability of storage capabilities at the edge servers \cite{EjderVR17}.    
	

\vspace*{.2cm}
\enablerLat {Parallel and Coded Computing}:
The computing task data can be distributed over multiple servers in different edge computing scenarios. For example, in a smart vehicle scenario where the navigation map data can be partly stored in several edge servers.  Parallel execution of computing tasks over multiple servers significantly impacts the efficiency and speed of task execution. Moreover, it eliminates the need to collect the full task dataset in a single entity. For example, \emph{partial offloading} can be performed where only a partition of the task is offloaded to where its required input data is available  \cite{Kaibin_MEC_survey}. The implementation of parallel computing depends on the correlation between the task partitions, i.e., only partitions that are not dependent on each other can be executed in parallel, whereas dependent tasks have to be executed sequentially. task dependency graph models and task partitioning \cite{Kaibin_MEC_survey,Chen_EdgeIoT18} are used to tackle the inter-dependency between the different task partitions.

 A challenge in realizing parallel computing, however, is the resulting high  inter-server communication load. Moreover, it suffers from the straggling effect, where a missing result from a single node delays the entire computation process. The concept of coded computing has shown to address both of these challenges \cite{CodedMR15}. Through exploiting the redundancy in the task partitions execution at different servers, coded multicast messages, e.g. via maximum distance separable (MDS) codes, can be used to deliver the results of the missing partitions simultaneously to multiple servers. This approach significantly reduced the amount of data that has to be communicated between the servers, at the expense of more redundant task executions at each server. Coded computing also helps in minimizing the overall computing latency through minimum latency codes. In conventional parallel computing task, each server executes a partition of the task and returns its result to the client. In this model, one delayed or failed partition will cause a delay or failure to the entire task. Alternatively, by generating redundant task data that are coded combinations of the original task data and executing these coded tasks, the result can be recovered by decoding the data from only a subset of the servers, eliminating the effect of a delayed or failed result. Optimizing the creation of the redundant coded tasks enables an inverse linear trade-off between the computing latency and computing load \cite{CodedComputing17}.

\subsubsection{High Reliability Enablers}
For \gls{mec} to fulfill its role and run applications on devices’ behalf, i.e. offloading the computing, it needs to be able to operate below stringent latency values, which are unachievable in traditional \gls{mcc} systems or too demanding to be run locally due to excessive computational and communication power

In this regard, to exploit both the high capacity of \gls{5g} mobile connections and the extensive computing capabilities located at the edge cloud, the concept of reliability is introduced with a two-fold interpretation:
In the first place, we find the classical notion of reliability related to error-robustness guarantees. As such, it allows to be tackled at different layers, including the reliability of the wireless link at the \gls{phy}.
Another fundamental notion of reliability, that has been widely adopted for wireless communications and standardization bodies as the \gls{3gpp}, is that of reliability understood as a probabilistic bound over the latency.

Understood in its most classical form, it is common that a toll in return for ensuring high reliability will have to be paid in the form additional/increased delays. 
For instance, at the \gls{phy} layer the use of parity, redundancy, and re-transmission will increase the latency. Also, in multi-user environments allocating multiple sources to a single user while clearly beneficial at an individual level, could potentially impact the experienced latency of the remaining users.

Next, we will set forth some of the enablers for both notions of reliability. 

\vspace*{.2cm}
\enablerRel{Multi-Connectivity}:
Compared to wired transmissions, in wireless environments temporary outages are common due to impairments in the \gls{sinr}.
These originate from, among others, stochasticity of the wireless channels, fluctuating levels of interference, or mobility of the \glspl{mue}. 
The term \glsreset{mxconn}\gls{mxconn}~\cite{Mxconn_Architectures_5G} encompasses several techniques developed with the overarching aim of enhancing effective data rates and the mobility robustness, i.e. the reliability, of wireless links. For that purpose, \gls{mxconn} exploits different forms of diversity to cut down on the number of failed handovers, dropped connections and, generally speaking, \glspl{rlf} that might cause service interruptions~\cite{Soret_ReliabilityLatencyThroughputTradeoffs_2014,Team_fettweis_mxconn_HowReliable_2018}. 

\gls{mxconn} solutions are classified as intra or inter frequency, i.e., depending on whether they operate using the same frequency or, otherwise, combine multiple carrier frequencies. Examples of the former include \gls{comp}~\cite{tech:CoMP_3gppTR36.819} transmissions and \glspl{sfn}~\cite{SFN_Seminal_Eriksson}. \gls{comp} involves a set of techniques that exploit rather than mitigate \gls{ici} to improve the performance at the cell edge. On performing joint processing, dynamic point selection (JP/DPS) or coordinated scheduling and beamforming (CS/CB) in the \glsunset{ul}\gls{ul}/\glsunset{dl}\gls{dl}, \glspl{bs} effectively operate as if assembled in a distributed multiple antenna system.
\glspl{sfn} embody a form of synchronous multicell transmission whereby various sources use the same time and frequency resource to non-coherently transmit signals to a receiver. The multiple received copies will be then constructively combined if their propagation delays are tightly bounded or, else, will induce \gls{isi}~\cite{Team_fettweis_mxconn_2017}. 

As for inter-frequency \gls{mxconn}, \gls{ca}~\cite{tech:CA_3gppTR36.823} and \gls{dc} are its most noteworthy examples. In \gls{ca} contiguous or non-contiguous component carriers, possibly allocated to several different \glspl{bs}, are combined and the scheduling and interference management orchestrated over these frequency bands aiming to enhance the resulting system's capacity. As for \gls{dc}, this framework provides solutions for inter-frequency, for \glspl{hetnet} scenarios, and for different wireless standards \gls{mxconn} so that a \gls{ue} will be simultaneously connected, respectively, in two different frequencies, to two different types of \glspl{bs} or two different wireless standards~\cite{Team_fettweis_mxconn_SFN_2015}. Recently, the idea of \gls{dc} for \gls{mmwave} and microwave bands has been proposed~\cite{Team_sundeep_mxconn_conf_2016,JHParkTWC:15} as an effective approach to facilitate cellular \gls{mmwave} \gls{ia}~\cite{Team_sundeep_IA_mag_2016} as well as   \gls{mmwave} handover~\cite{Team_sundeep_DC_handover_2017}. In like manner, \gls{mmwave} and sub 6~GHz \gls{dc} can team together to augment the reliability of the \gls{mmwave} working as fallback to compensate eventual \gls{mmwave} channel vulnerability, e.g. to blocking events. Finally, the benefits of integrating communication interface diversity for reliability purposes are also studied in~\cite{team_popovski_latencyUR_2016} in the context of \gls{mtc}. 

\gls{sfn} operation is proposed in use case 6 detailed in Section \ref{usecaseVR1}. The goal is to protect against \gls{mmwave} channel intermittence by increasing the rate of those links between the \glspl{mmap} and the \glspl{vrp} that, otherwise, would jeopardize the immersive experience.

\vspace*{.2cm}
\enablerRel{Task Replication}: While \gls{mxconn} can boost the reliability in the presence of channel fluctuations, it requires coordination between the different servers that are connected to the end user. However, when coordination is not possible, reliability can still be enhanced through the task replication. Similar to packet replication in data communication, a user can offload a computing task to multiple servers that are not connected to each other and receive the result from whichever has the result ready first. This mechanism provides more guarantees of task execution, at the expense of reduced system capacity, due to the under-utilization of computing servers. One realization of this concept is proposed in \cite{tail_at_scale}, namely, \emph{hedged requests}, is when the user sends one replica of the task to the server that is believed to be most suitable, then follows by sending another replica to an additional server after some delay. Completion pending remaining requests are canceled once a result is received from any server. 

While task replication is can be efficient in ensuring the reliability in in the case of channel dynamics, it incurs significant additional load. To combat this, one can offload the task to an additional server only when the delay from the first server exceeds a certain threshold \cite{tail_at_scale} This approach is investigated in \cite{ElbambyEurasip2018}. Therein, it shown that imposing such condition can significantly curb the latency variability without inducing much additional load.

\vspace*{.2cm}
\enablerRel{Federated Machine Learning}:
While performing \gls{ml} inference at the network edge yields low latency, distributed training of their \gls{ml} models across different edge nodes improves the inference reliability. To be specific, each learning agent optimizes its \gls{ml} model during the training phase so as to maximize the inference accuracy over locally available training data. The measured inference accuracy at the training phase is however not always identical to the inference accuracy at the test phase, primarily because of unseen training data samples. This accuracy gap is known as the \emph{generalization error} that measures the inference reliability under unseen data samples~\cite{Bosquet:2004}. A straightforward way to reduce the generalization error is exchanging training data samples among edge nodes. Data exchange, however, incurs extra communication and computation cost, and may not be available for user-generated private data. To address this problem, \gls{fl} has recently been proposed~\cite{pap:jakub16,Brendan17}, in which edge nodes exchange and aggregate their local \gls{ml} models, thereby preserving data privacy, avoiding extra computation, and reducing communication overhead when \gls{ml} model sizes are sufficiently smaller than data sizes.

\Gls{fl} is still a nascent field of research, calling for co-designing communication, computation, and \gls{ml} architectures~\cite{Park:2018aa,KaibinIntEdge18}. For instance, the original \gls{fl} algorithm has the communication payload size being proportional to the \gls{ml} model sizes, and thus cannot deal with deep neural network models. Proper model compression and parameter quantization techniques are thus needed, while trading the increased communication efficiency off against the reduced accuracy. Furthermore, the server in current \gls{fl} algorithms simply aggregates uploaded local models, although it has higher computation resources compared to the edge devices. Along with these \gls{fl} architectures, computing task offloading, task scheduling, and resource allocations should be jointly optimized towards achieving reliability under uncertainties on \gls{mec} operations, including unseen data samples, channel fluctuations, and time-varying communication and computation resources.

\vspace*{.2cm}
\enablerRel{Extreme Event Control}:
As mentioned previously, one reliability notion is the probability of violation or failure over a latency bound, which can be mathematically expressed as $\Pr(\mbox{Latency}>L_{\rm bound})$.  This probability ranges from $10^{-3}$ to $10^{-9}$, depending on the mission-critical application in 5G networks \cite{MehdiURLLC:18}. To meet the ultra-reliability requirements, we should focus on the extreme events with very low occurrence probabilities.
However, in classical communication systems, the designed approaches are based on the expected metrics, e.g., average rate and average latency, in which the random event realizations with higher probability  distribution function (PDF) values dominate the system performance. In other words, the conventional average-based approaches are inadequate for enhancing reliability performance, and instead we need to take into account the metrics or statistics, which are related to or affect the extreme events, such as
\begin{itemize}
\item
worst-case measurement, e.g., largest latency in the network,
\item
tail/decay behavior of the complementary cumulative distribution function (CCDF),
\item
very low bound violation probability,
\item  
threshold deviation and its higher-order statistics, e.g., variance,
 \end{itemize}
while designing the URLLC-enabled MEC systems. 
To analytically analyze these metrics and statistics,  extreme value theory (EVT) \cite{EVT:Cole,EVT:Han} is a useful methodology for mathematical characterization and, thus, provides a powerful framework for extreme event control. Let us introduce the fundamental theorems in EVT as follows, which characterize the aforementioned metrics and their statistics.

\begin{theorem}[{\bf Fisher--Tippett--Gnedenko theorem \cite{EVT:Cole}}]\label{Thm: GEV}
We consider $n$  independent and identically distributed ({\it i.i.d.})~samples from a random variable $X$, i.e., $X_1,\cdots,X_n\overset{i.i.d.}{\sim}X$ and define $Z_n\coloneqq\max\{X_1,\cdots,X_n\}$. If $Z_n$ converges to a non-degenerate distribution as $n\to\infty$, we can approximate the limit as a generalized extreme value (GEV) distribution which is characterized by a location parameter $\mu\in\mathbb{R}$, a scale parameter $\sigma >0$, and a shape parameter $\xi \in\mathbb{R}$. 
\end{theorem}
Among them, the shape parameter governs the GEV distributions' tail behaviors \cite{EVT:Han}, which are sorted into three types depending on the value of $\xi$.
\begin{enumerate}
\item
When $\xi>0$,   the GEV distribution has a  \emph{heavy-tailed} CCDF which is more weighted than an exponential function.
\item
When $\xi=0$, the GEV distribution has a \emph{light tail}, in which the CCDF has a thinner tail than an exponential function. 
 \item
When $\xi<0$, the GEV distribution is \emph{short-tailed}. That is, the CCDF has a finite upper endpoint at $z=\mu-\sigma/\xi$.
\end{enumerate}
When $\xi\geq0$, the upper endpoint of the CCDF approaches infinity.

\begin{theorem}[{\bf von Mises conditions \cite{EVT:Han}}]\label{Thm: von mises}
In Theorem \ref{Thm: GEV}, the characteristic parameters $(\mu,\sigma,\xi)$ of the approximated GEV distribution can be asymptotically found as per
$\mu=\lim\limits_{n\to\infty} F_X^{-1}(1-1/n),$
$\sigma=\lim\limits_{n\to\infty} \frac{1}{nf_X(F_X^{-1}(1-1/n))},$
and $\xi=-1-\lim\limits_{x\to\infty}\frac{[1-F_X(x)]f^{'}_X(x)}{[f_X(x)]^2}.$
\end{theorem}
\begin{theorem}[{\bf Pickands--Balkema--de Haan theorem \cite{EVT:Cole}}]\label{Thm: Pareto}
Consider the random variable $X$ in Theorem \ref{Thm: GEV} and a threshold $d$. As $d\to F^{-1}_{X}(1)$,  the CCDF of the excess value $Y|_{X>d}=X-d>0$ can be approximated as a generalized Pareto distribution (GPD)
whose  mean and variance are  $\tilde{\sigma}/(1-\xi)$ and $\frac{\tilde{\sigma}^2}{(1-\xi)^2(1-2\xi)}$, respectively.
\end{theorem}

Analogously to the GEV distribution, the GPD is characterized by a scale parameter $\tilde{\sigma}>0$ and a shape parameter $\xi\in\mathbb{R}$. In Theorems \ref{Thm: GEV} and \ref{Thm: Pareto},  $\xi$  is identical while $\sigma=\tilde{\sigma}+\xi(\mu-d)$. Note that Theorems \ref{Thm: GEV} and \ref{Thm: von mises} provide a way to characterize the worst-case metric and its tail behavior, whereas Theorem \ref{Thm: Pareto} is directly related to the  bound violation and its statistics. Since the characteristic parameters of the GEV distribution and GPD are identical or related, the results of these three theorems are complementary to one another.

Nevertheless, some tradeoffs and dilemmas exist when we apply the results of EVT and estimate the characteristic parameters. For example, we need to trade off data availability, which affects the performance, convergence speed, and estimation accuracy.
Specifically, given $N$ {\it i.i.d.}~realizations of $X$ (i.e., $N/n$ realizations of $Z_n$), larger $n$ theoretically gives the better approximation of the GEV distribution but slows down the convergence of parameter estimations due to the less availability of data samples of $Z_n$.
The similar tradeoff between high threshold $d$ and availability of threshold-exceeding data can be found from Theorem \ref{Thm: Pareto}.
Additionally, if the distribution of $X$, e.g., delay of a single user, is unknown beforehand, this agnostic makes Theorem \ref{Thm: von mises} difficult to characterize the network-wide largest delay.
Fortunately, thanks to the mature development in the ML field, the aforementioned issues can be tackled by using the ML approaches, in which unsupervised learning provides a way to infer a mathematical expression of the unknown distribution, while the lack of available data is addressed in an FL manner by aggregating and averaging the estimated characteristic parameters of all distributed devices.

\section{Applications and Use cases}

In this section, we elaborate on some of the prospective services and applications for whom offloading their computing tasks to the edge significantly improves their performance in terms of latency and reliability. In particular, we focus on two scenarios where offloading task computing to the network edge will be beneficial: 1) when end users have limited computing capabilities, e.g., \gls{vr} \glspl{hmd}); and 2) when end users have sufficient computing and energy resources, but are accessible only to a fraction of the entire information for the computation input, e.g., vehicular edge computing scenarios. We follow by presenting different edge computing use cases in which the  \gls{urllc} enablers are utilized.  

\subsection{Edge Computing Applications}
\subsubsection{Extended Reality}\label{usecaseVR}
\Gls{xr} is an umbrella term that covers all virtual or combined real-virtual environments, including \gls{vr}, \gls{ar} and \gls{mr}. These environments differ in the nature of the content a user sees or interacts with. While \gls{vr} describes environments where users are fully immerse in a virtual world, \gls{ar} refers to the view of a virtual environment that is merged or supplemented by elements or inputs from the real-world. \gls{ar} can be categorized as a special case of the more general \gls{mr}, which refers to the environments that mixes together real and virtual elements that can interact with each other. 

\gls{xr} is anticipated to be one of the leading applications to leverage edge computing. Providing high quality \gls{xr} experience comes with high computation resource demand. At the same time, \gls{xr} applications are highly sensitive to delay. Typically, a maximum \gls{e2e} delay, also known as \gls{mtp} delay, of 15-20 milliseconds can be tolerated in \gls{vr}. Higher delay values trigger what is known as motion sickness, resulting from a visual-motor sensory conflict. This makes it unrealistic to rely on remote cloud servers for processing. On the other hand, Processing \gls{xr} locally on the user device has several complications. First, \gls{xr} devices, such as \glspl{hmd} and smartphones are often equipped with limited compute capabilities. This limitation is due to the device size, manufacturing cost, as well as to limit the heat generated from powering the device. Second, running applications on different types of devices, with different hardware, operating systems, and platforms is a challenging task. For these reasons, existing standalone \gls{xr} devices often provide limited content quality. Standalone~\gls{vr} headsets operate with reduced frame resolution and frame rate~\cite{VR_Network_Mag}, whereas AR headsets such as Microsoft HoloLens restrict the amount of renderable polygons~\cite{AR_EC}. 

For these reasons, the success of \gls{xr} requires providing high computation and storage resources close to the end users. In this regard, edge computing is an intuitive solution to provide such services~\cite{conf:Elbamby2018-WCNC}. Today's most powerful \gls{vr} headsets rely on edge computers to perform sophisticated rendering. However, wired connections are still used between the headsets and the edge servers, due to the high rate requirement of \gls{vr} applications. This limits the mobility and convenience of \gls{vr} users and hence decrease the \gls{qoe}.

The need for a better \gls{xr} \gls{qoe} and the advancement in wireless communication capabilities motivate the development of wireless \gls{xr} systems that incorporate powerful edge computers and high capacity wireless links~\cite{ABIQualcommVR:17, ParkGC:18,OsvaldoAR:17,conf:Elbamby2018-WCNC,PerfectoVRStreaming}. \Gls{mmwave} communication can provide large spectrum and high data rates, making it a solid candidate for wireless \gls{xr}. Moreover, the directionality of \gls{mmwave} links allow for leveraging multi-user transmission techniques such as multicasting and broadcasting to deliver common and correlated content to multiple users in a way that minimizes the communication delay.  However, directional \gls{mmwave} links suffer outages due to signal blockage. This affects the link signal quality and increases the channel variability, and hence decreases the link reliability. \gls{mxconn} can be a viable solution to provide robust \gls{mmwave} communication. Using \gls{mxconn}, an \gls{xr} user maintains multiple simultaneous communication links with multiple servers.  

\vspace*{0.5cm}
\subsubsection{Vehicular Edge Computing and V2X/V2V for ADAS:}\label{usecaseV2V}
Future autonomous driving vehicles comprised as nodes of the \gls{iov}, a larger mobility network which can be considered as an extended application of the \gls{iot} to \glspl{its}\cite{WinWinMAG_Huawei_2011}, will operate as hubs integrating multiple technologies and consuming and producing massive volumes of data~\cite{RoadmapFutuAuto_McKinsey_2014}. 
The \glspl{adas} to be equipped in these vehicles, especially those pertaining to the area of traffic safety, heavily depend on reliable and instantaneous decision-making processes that hinge on inputs from multiple sensory data sources, including \gls{lidar}, automotive radar, image processing, computer vision, etc.~\cite{conf:Perfecto2017-EuCNC}. 
As an example, we can think of successful object identification from \gls{lidar} point clouds or speed and trajectory prediction for dynamic objects moving within a vehicle's vicinity. 
Hereof, it is essential that these vehicles are equipped with powerful computing and processing capabilities to swiftly handle high data volumes rather than solely relying on cloud services that, in the above example, may classify the objects or predict trajectories from raw data with higher accuracy but, possibly, incurring to do so in unacceptable delays. 
Moreover, for next-generation \gls{adas} it is envisaged that vehicles will communicate with each other as well as with an increasingly intelligent roadway infrastructure through the use of \gls{v2x} and \gls{v2v} communications, ultimately exploiting high capacity \gls{mmwave} links~\cite{jnl:Perfecto2017-JSAC,conf:Perfecto2017-interplay}.
Consequently, the cumbersome volume of locally generated data could be exacerbated by the acquisition of data from both the environment and from surrounding vehicles.

Indeed, vehicular edge computing will play a pivotal role to support delay-sensitive as well as future emerging multimedia-rich applications in vehicular networks, which is buttressed by the growing body of literature devoted to the area of content-centric applications of vehicular \gls{mec} \cite{VehicularEdge_ContentCentric_2016,VehicularEdge_PredOffloading_2017,VehicularEdge_ContentDelivery_2018,VehicularEdge_BigDataEE_2018} that are frequently combined with \gls{ml} to provide reliability as edge analytics~\cite{ITS_DL_MEC_Walid_2019}, to leverage huge volumes of information~\cite{VehicularEdge_BigDataEE_2018} or to provide an integrated framework for dynamic
orchestration of networking, caching, and computing resources in next generation vehicular networks~\cite{Vehicular_DL_Caching+Computing_2018}.

Being not nearly as tightly constrained by size or by the access to a power supply as their counterpart \gls{iot} devices or smartphones, the computational and storage capabilities in vehicular terminals could allow them to run locally or collaboratively, using vehicles as the infrastructures for communication and computation as proposed in~\cite{Vehicular_VehAsInfrastructures_2016}, resource-hungry applications\footnote{However, the longer product's life-span in the automotive industry, according to the US Department of Transportation as of 2018 the average age of on-the-road vehicles is over 11 years~\cite{CarLifeSpan_2018}, could quickly turn onboard \gls{cpu}/\gls{gpu} processing capabilities obsolete.}. 
In this regard, provided that computing and processing capabilities may not be the limiting factor, a second advantage of running these applications in the network edge is substantiated by the availability of data collected from multiple vehicles in edge servers. 
Access to this information raw or preprocessed can augment individual vehicles' situational awareness by extending their own sensing range. Resorting to edge contents can thus provide \emph{a bigger picture} at acceptable delays.

The later idea is exemplified in the third usecase in upcoming Section \ref{usecaseEVTFL-V2V} where the information from different vehicles is combined in the network edge following \gls{fl} principles and used to refine a global model for transmission queue length distribution for the purpose of providing ultra-reliable low-latency \gls{v2v} communications.

\subsection{Use Cases}
Next, we present different case studies in which the \gls{urllc} enablers are utilized in edge computing settings.
%
%
%

\begin{figure}[t!]
\centering
\subfigure[Data split.]{
	\includegraphics[width=.45\columnwidth]{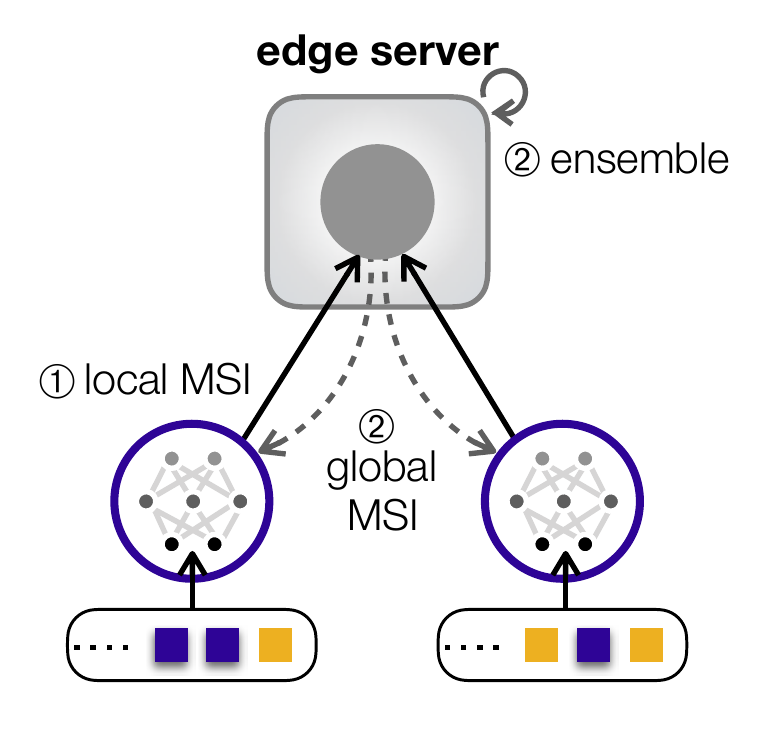}
	\label{Fig:split_helper_device}
}
\subfigure[Model split.]{
	\includegraphics[width=.45\columnwidth]{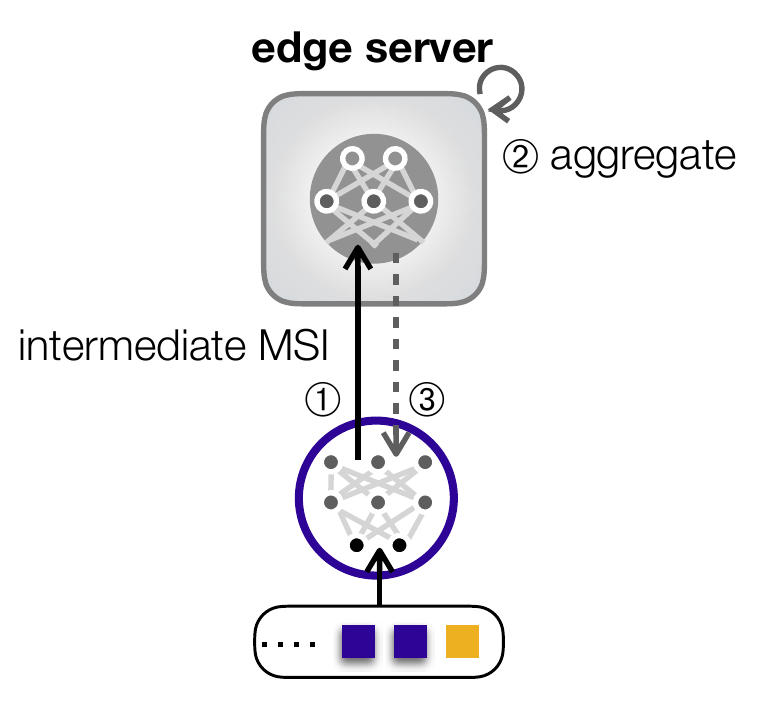}
	\label{Fig:split_model}
}
\caption{Edge \gls{ml} architectural splits: (a) data split and (b) model split.}
\label{Fig:arch_split}
\end{figure}

\begin{figure}[t!]
\centering
    \subfigure[Communication cost.]{
    	\includegraphics[width=.8\columnwidth]{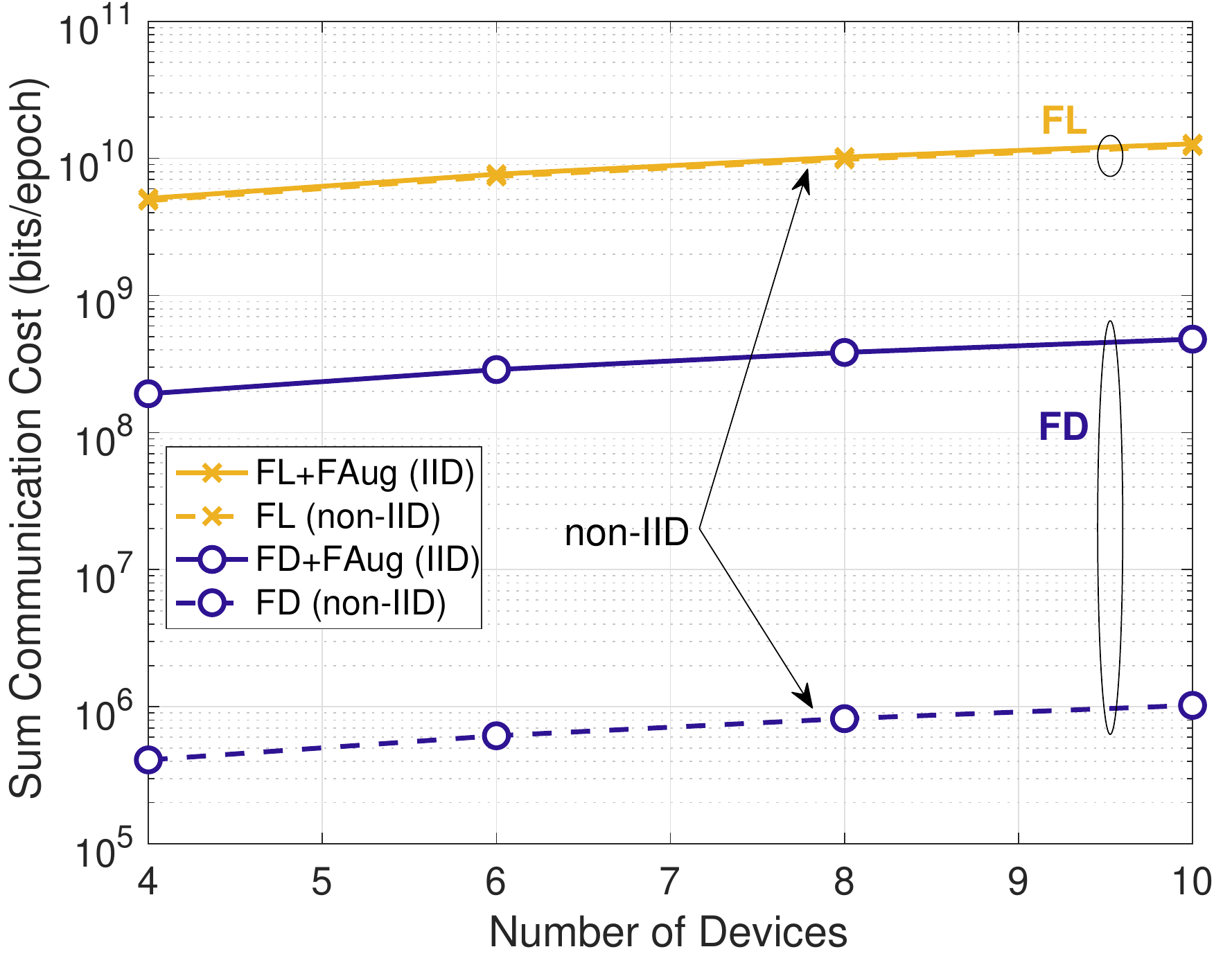}
    	\label{Fig:FD_comm}
    }
    \subfigure[Test accuracy.]{
    	\includegraphics[width=.8\columnwidth]{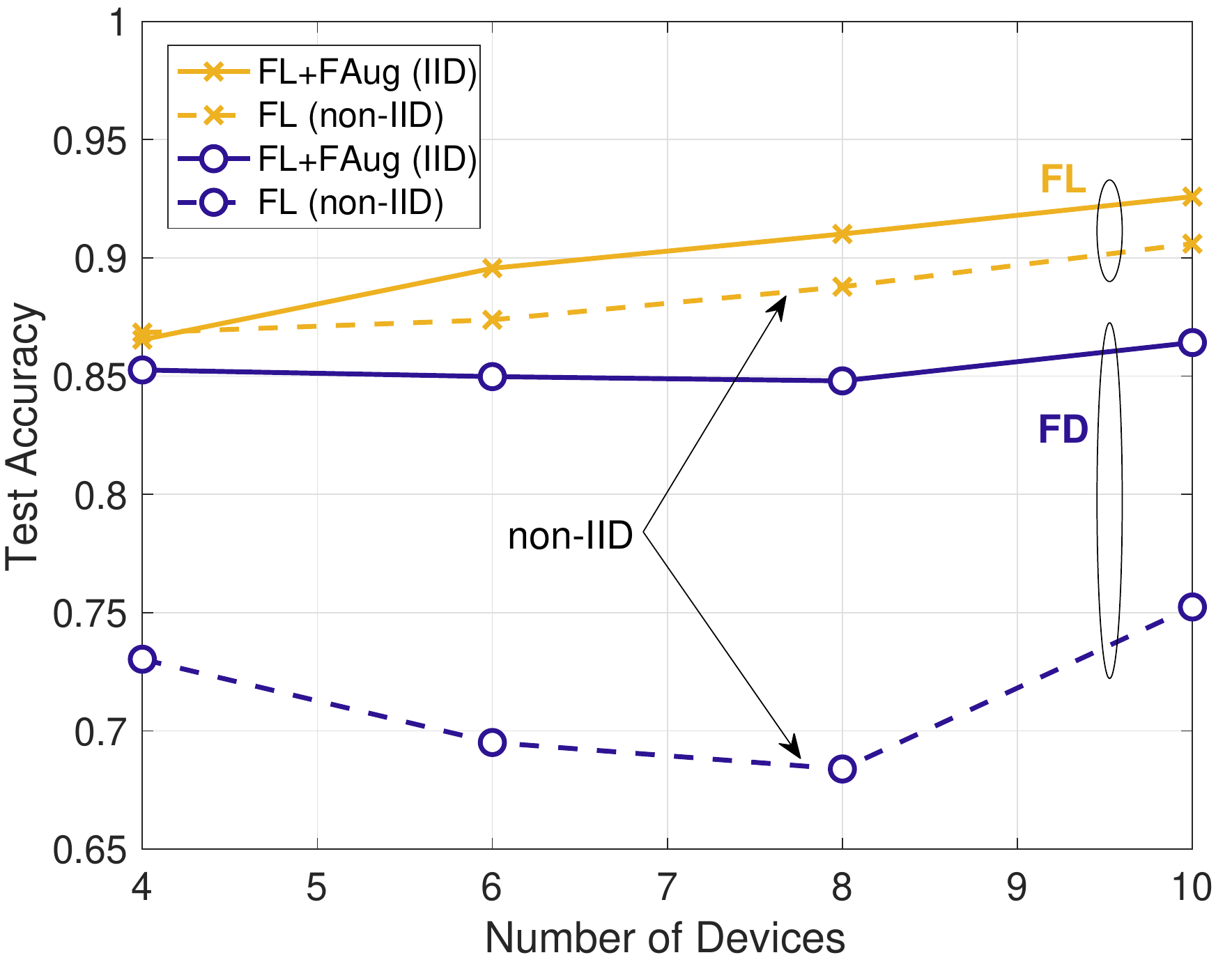}
    	\label{Fig:FD}
    }
\caption{Communication cost and inference accuracy of federated learning~(\gls{fl}) and federated distillation (FD) with or without federated augmentation (FAug) in the MNIST classification problem, where each device stores a 5-layer convolutional neural network (CNN). For FAug, the conditional generative adversarial network (GAN) consists of a 4-layer generator \gls{nn} and another 4-layer discriminator \gls{nn}.
}\label{Fig:FDFL}
\end{figure}

\vspace*{0.2cm}
\label{usecaseDistML}
\usecase{Edge Computing for Federated Machine Learning}: As addressed in Sect. 3.2.1 and 3.2.2, edge \gls{ml} is envisaged to be a key enabler for \gls{urllc}, in which both inference and training processes of \gls{ml} models, e.g., \glspl{nn}, are pushed down to the network edge~\cite{Park:2018aa}. This direction of edge\gls{ml} has been fueled by \gls{fl}~\cite{pap:jakub16,Brendan17,KimWCL:18,Amiri:2019,Ha:19,Wang:2018aa} under a \textit{data split} architecture (see Fig.~\ref{Fig:split_helper_device}), where edge devices collectively train local models with their own user-generated data via a coordinating edge server that aggregates locally computed model updates, referred to as \gls{msi}. The \gls{mec} framework can further improve \gls{fl} by its co-design with training architectures and algorithms. In view of this, on the one hand, each edge device is able to optimize the \gls{msi} type depending on the \gls{nn} model size and channel quality. As done in \gls{fl}, one can exchange the model parameter \gls{msi} whose payload size is proportional to the model size, which is not feasible for deep \glspl{nn} under poor channel conditions. Alternatively, one can exchange model output MSI whose payload size is independent of the model size, referred to as federated distillation (FD)~\cite{Jeong:18}. As shown in Fig.~\ref{Fig:FD_comm}, this fundamentally results in FD's incomparably smaller communication payload per MSI exchange than FL, and can thereby better cope with poor channel conditions.

On the other hand, the edge server can assist in the training process by exploiting its extra computation and communication resources. A compelling example is to rectify the non-IID training dataset incurred by the user-generated nature of data, wherein entirely un-correlated (non-identical) and/or too similar (non-independent) data samples across devices negate the benefit of distributed training~\cite{GoodfellowBook:16}. To this end, in federated augmentation (FAug)~\cite{Jeong:18}, the edge server first collects few seed samples from edge devices, and oversamples them (e.g., via Google's image search for visual data) through its fast connection to the Internet. Then, the edge server can utilize its high computing power for training a generative model (e.g., conditional generative adversarial network (GAN)~\cite{CondGAN14}). Downloading the trained generator empowers each device to locally augment deficient data samples until reaching an IID training dataset. With FAug, both FL and FD yield higher test accuracy as shown in Fig.~\ref{Fig:FD}, at the cost of slight increase in communication cost as illustrated in Fig.~\ref{Fig:FD_comm}.

Lastly, a very deep \gls{nn} (e.g., Inception V4 \gls{nn} model consuming 44.3 GB~\cite{Wang:2018:inception}) cannot fit into a single device's memory, and has to be partitioned into multiple segments stored across edge devices and server, i.e., \textit{model split} (see Fig.\ref{Fig:split_model}). Here, the model's local and offloaded computations should be orchestrated over wireless links, by optimizing the partitioning strategy based on the \gls{nn}'s topology and constituent layers. This calls for a novel \gls{mec} framework that takes into account not only communication and computation resources but also \gls{nn} forward and backward propagation dynamics intertwined with channel dynamics.

\begin{figure*}[t]
\centering
\includegraphics[width=.75\textwidth]{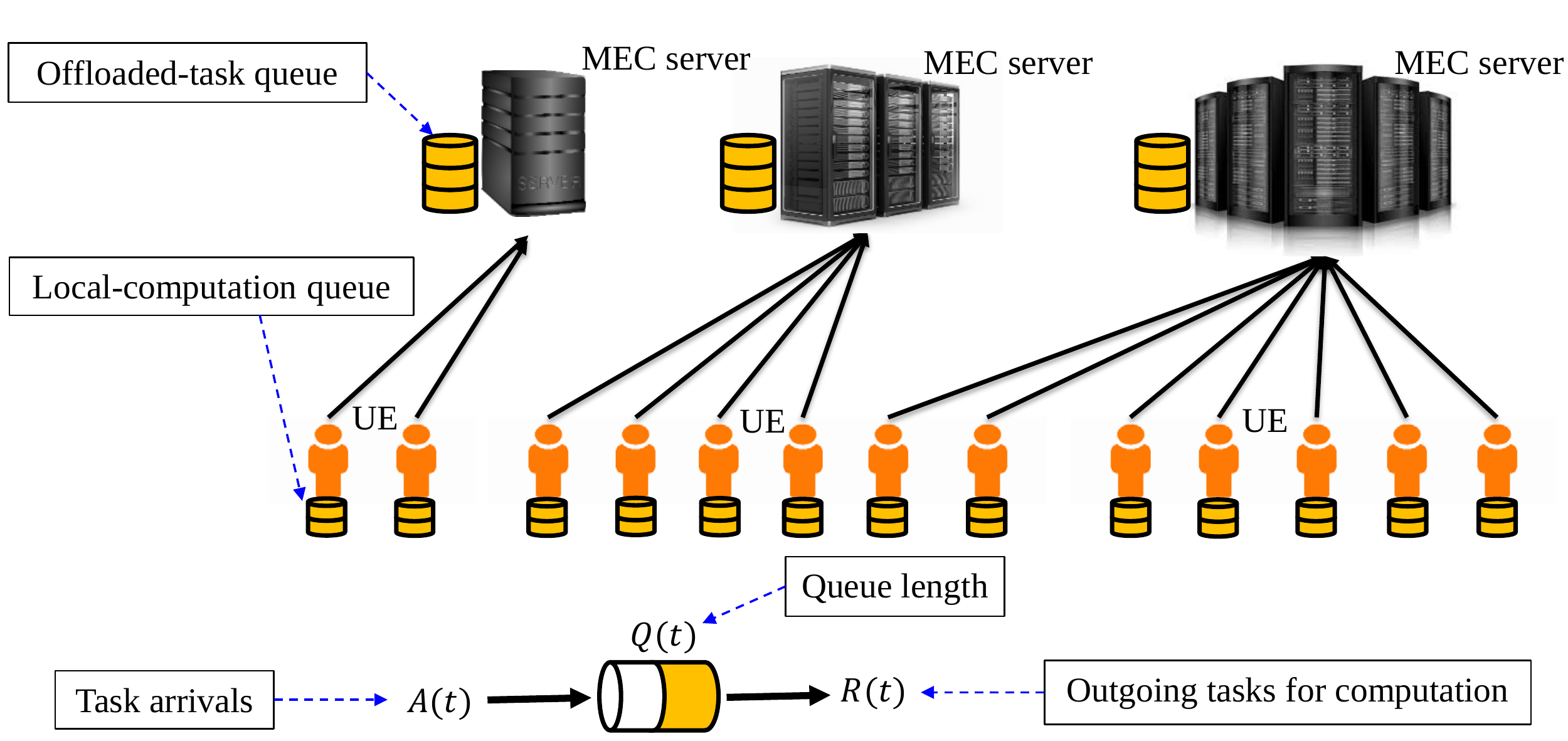}
	\caption{Extreme Event-Controlled MEC  architecture.}
		\label{Fig: TCOM system model}
\end{figure*}

\vspace*{0.3cm}
\label{usecaseEVT}
\usecase{Extreme Event-Controlled MEC}:
For the extreme event-controlling computation and communication co-design in \cite{CFLiu_MECTaskOffloading_Globecom,CFLiu_URLLC_MECTaskOffloading_TCOM}, we studied a multi-user \gls{mec} scenario as shown in Fig.~\ref{Fig: TCOM system model}, in which multiple \gls{mec} servers with different computation capabilities are deployed. In this setting, the \gls{ue} manages its local resource (i.e., total power budget) for computation and communication, i.e., task offloading, while the \gls{mec} server schedules its computational resources for the \glspl{ue}' offloaded tasks. Herein, we consider the length of the task queue as a latency measurement since queuing latency can be reflected by the queue length. For the reliability concerns, we are concerned about the bound violation probability
and higher-order statistics of threshold deviation as highlighted in high reliability enabler 4. In this regard, we first impose a constraint on the queue length\footnote{The notation $Q$ generalizes the lengths of all task queues at the \glspl{ue} and \gls{mec} servers.} bound violation probability as
\begin{figure*}[t!]
\centering
\subfigure[]{
	\includegraphics[width=.32\textwidth]{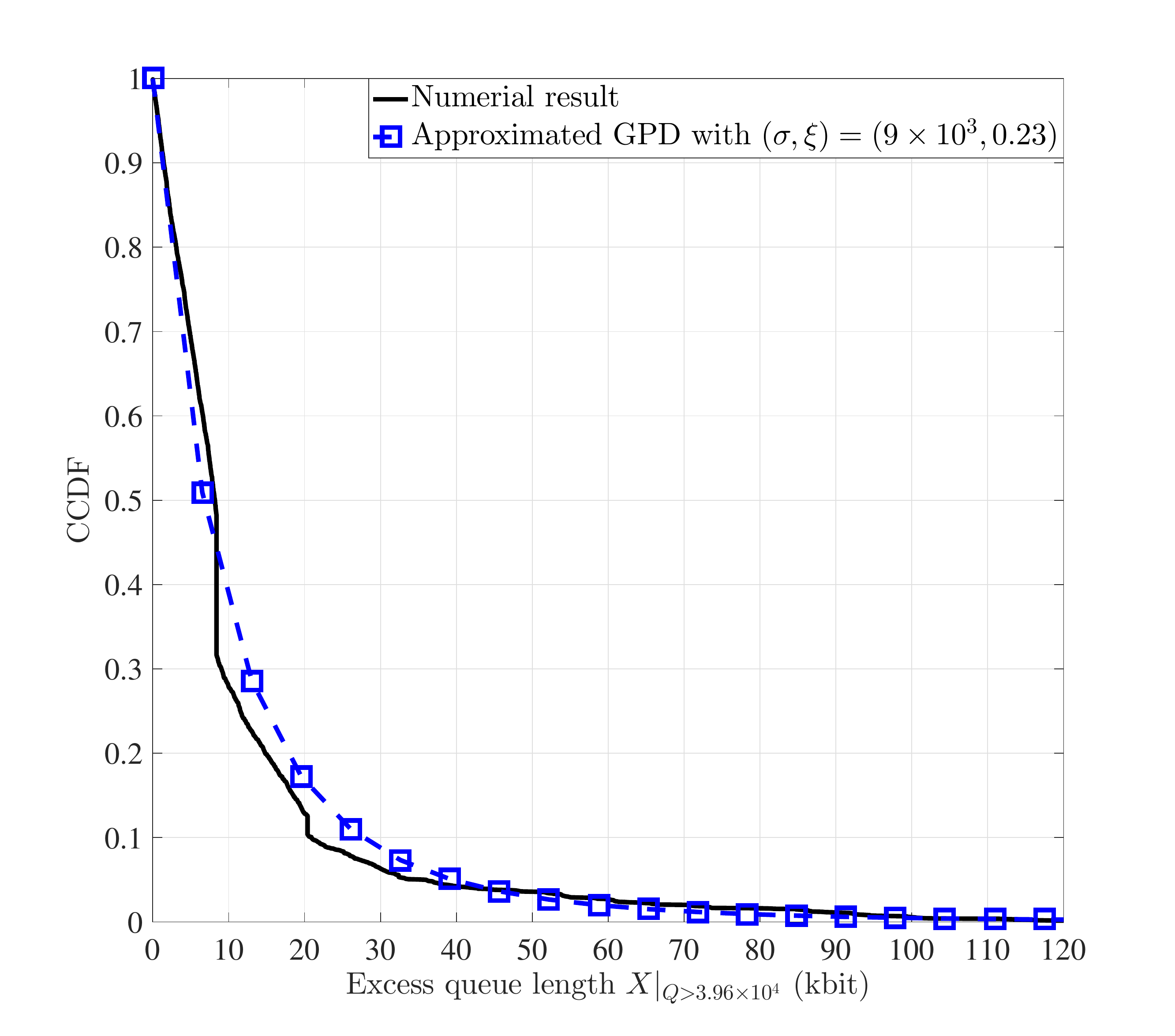}
	\label{Fig:TCOM_results_1}
}
\subfigure[]{
    \includegraphics[width=.32\textwidth]{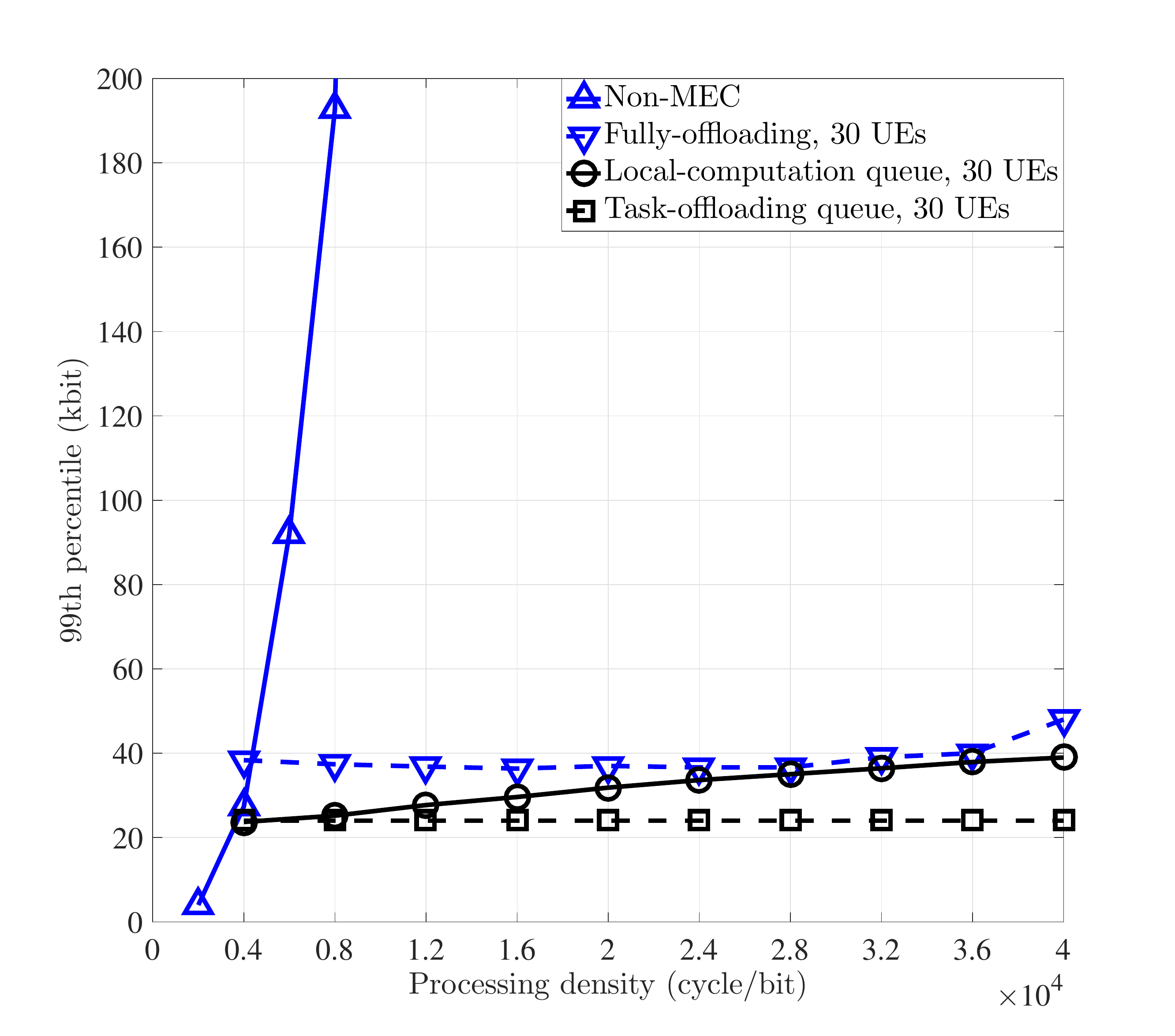}
	\label{Fig:TCOM_results_2}
}
\subfigure[]{
    \includegraphics[width=.32\textwidth]{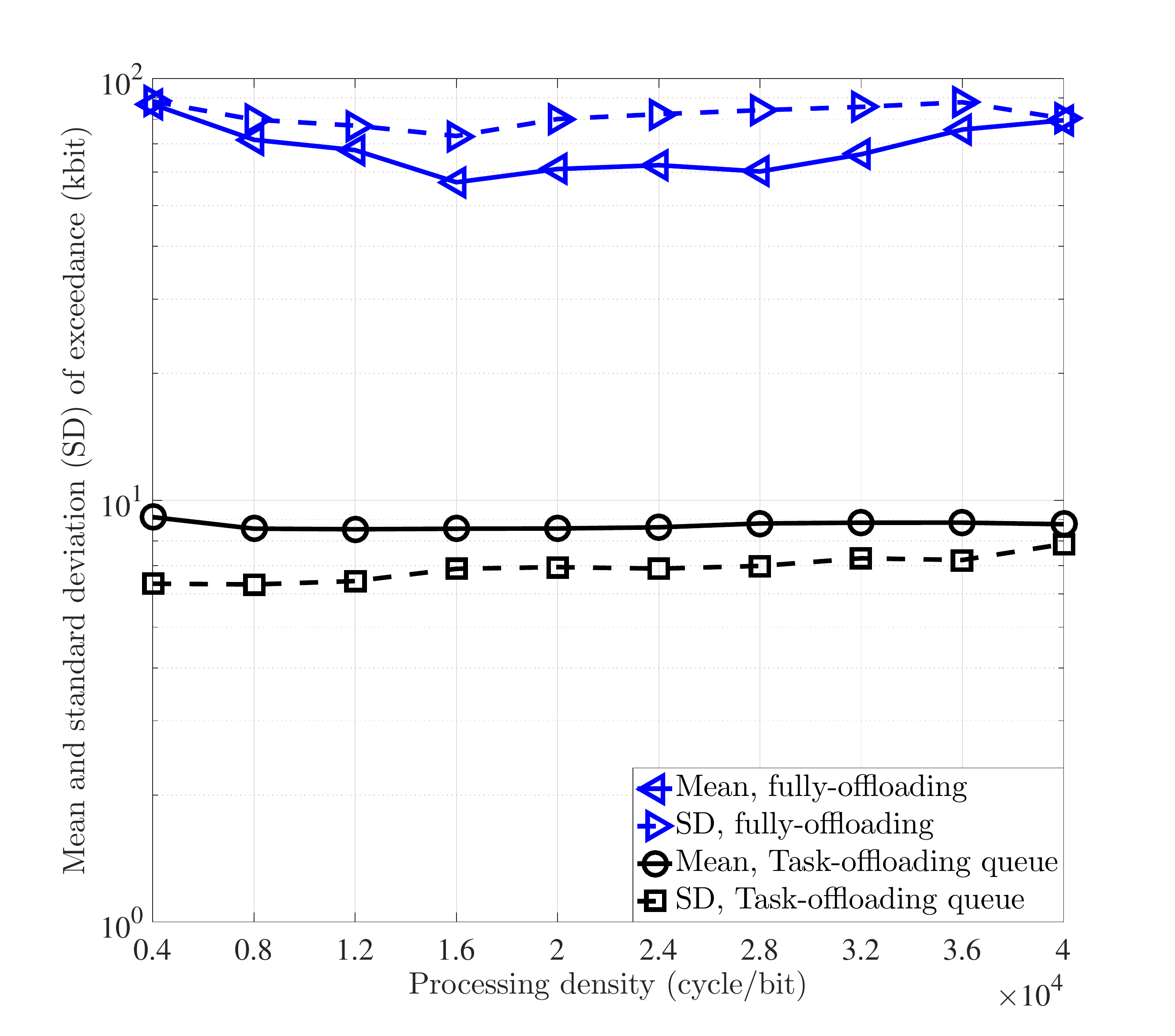}
    \label{Fig:TCOM_results_3}
}
\caption{(a) Tail distributions of the excess queue length and the approximated GPD of exceedances, (b) 99th percentile of the queue length, and (c) mean and standard deviation of exceedances over the 99th percentile queue length, versus  processing density.}
\label{Fig:TCOM_results}
\end{figure*}

%
%
%
\begin{align}
 &\lim\limits_{T\to\infty}\frac{1}{T}\sum\limits_{t=1}^{T}\Pr\big(Q(t)> d \big)\leq \epsilon\ll 1.\label{Eq: Violation-Loc-Prob}
\end{align}
Here, $d$ and $ \epsilon$  are the given bound and tolerable violation probability.
Let us further focus on the excess value over the bound $d$, which is denoted by  $X(t)|_{Q(t)> d}=Q(t)- d>0$. 
By applying Theorem \ref{Thm: Pareto},  we approximate the exceedances as a GPD with the characteristic parameters   $(\tilde{\sigma},\xi)$. The mean and variance are
$\mathbb{E}\big[X(t)|Q(t)> d \big] \approx\frac{\tilde{\sigma}}{1-\xi}$
and 
$\mbox{Var}\big(X(t) |Q(t)> d\big) \approx\frac{\tilde{\sigma}^2}{(1-\xi)^2(1-2\xi)}$,
respectively. We can find that the smaller  $\tilde{\sigma}$ and $\xi$ are, the smaller the mean value and variance. Since the approximated GPD is just characterized by the  scale and  shape parameters, we impose thresholds on these two parameters, i.e.,  $\tilde{\sigma}\leq \tilde{\sigma}^{\rm th}$ and $\xi\leq \xi^{\rm th}$. 
Subsequently, applying the two parameter thresholds and $\mbox{Var}(X)=\mathbb{E}[(X)^2]-\mathbb{E}[X]^2$, we consider the conditional constraints on the  mean and second moment of the excess queue length
\begin{align}
&\lim\limits_{T\to\infty}\frac{1}{T}\sum\limits_{t=1}^{T}\mathbb{E}\big[X (t)|Q(t)\!>\! d \big]\!\leq\! \frac{\tilde{\sigma}^{\rm th}}{1-\xi^{\rm th}},\label{Eq: GPD-Loc-mean}
\\&\lim\limits_{T\to\infty}\frac{1}{T}\sum\limits_{t=1}^{T}\mathbb{E}\big[[X(t)]^2 |Q(t)\!>\! d\big]\!\leq\!\frac{2\big(\tilde{\sigma}^{\rm th}\big)^2}{\big(1-\xi^{\rm th}\big)\big(1-2\xi^{\rm  th}\big)}.\label{Eq: GPD-Loc-var}
\end{align}
%
%
%
%
%
%
%
%
%
%
%
Taking into account the above three requirements for the extreme events,
 we trade off the UE's
computation power 
and communication power in the extreme event-controlling computation and communication co-design.

The effectiveness of characterizing threshold deviation by the Pickands--Balkema--de Haan theorem, i.e., Theorem~\ref{Thm: Pareto},  is verified in Fig.~\ref{Fig:TCOM_results_1}. Therein, $\Pr(Q>d)=3.4\times 10^{-3}$ with $d=3.96\times 10^{4}$. Additionally, in contrast with the schemes without edge computing and without local computation capability, the extreme event-controlling approach achieves the better performance, in terms of the extreme event-related metrics shown in  Fig.~\ref{Fig:TCOM_results_2} and Fig.~\ref{Fig:TCOM_results_3}, in the considered MEC system. 
%
%
%
%
%
%
%

%
%
%
%

\begin{figure*}[t]
    \centering
	\includegraphics[width=0.8\textwidth]{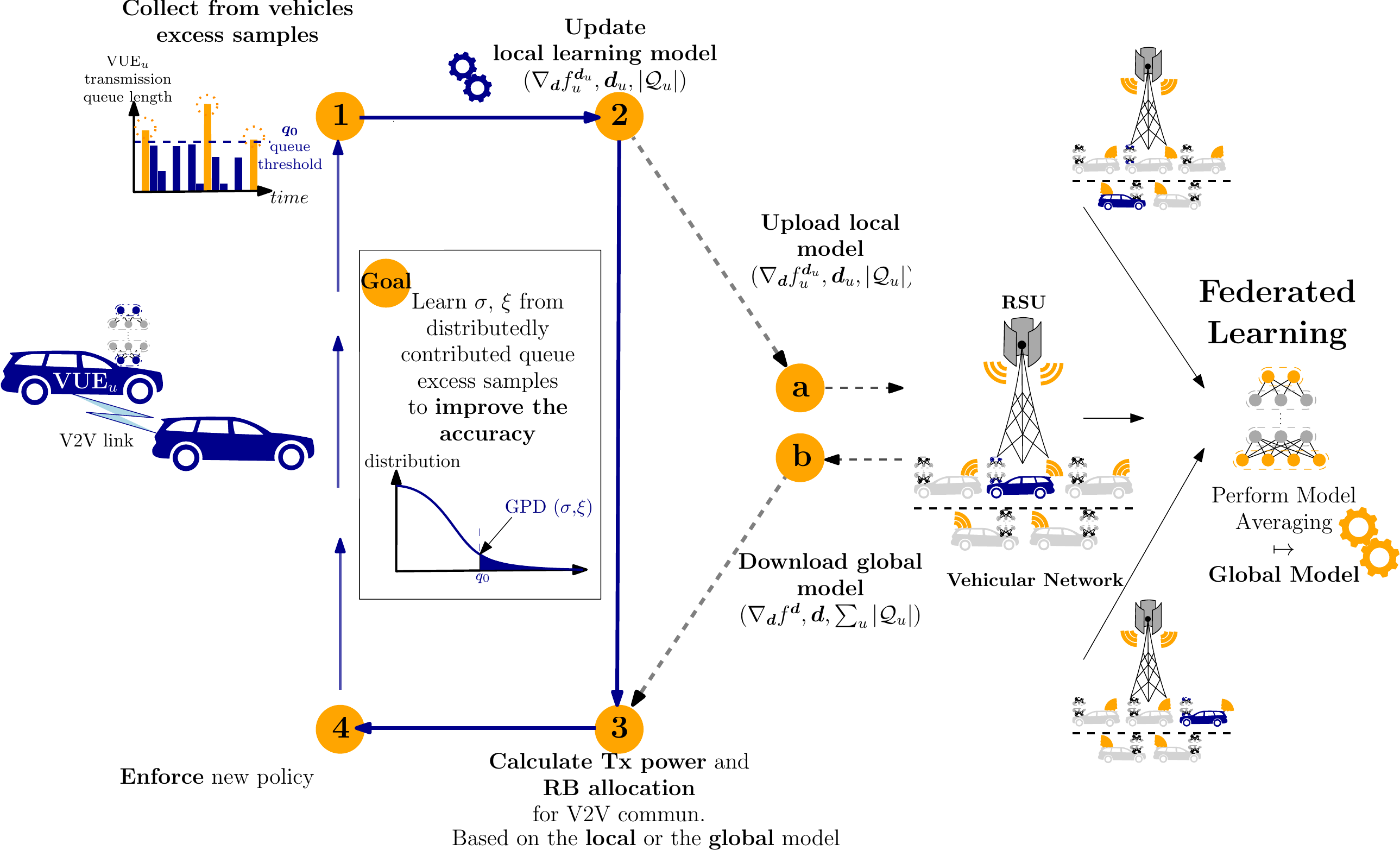}
 	\caption{Operational structure of EVT parametric FL ($\ASYNC$).}
 	\label{fig:v2v_fl_mle}
\end{figure*}

\vspace*{0.35cm}
\label{usecaseEVTFL-V2V}
\usecase{EVT/FL Ultra-Reliable Low-Latency V2V Communication}:
The idea of how to combine \gls{evt} and \gls{fl} to enable \gls{urllc} in vehicular communication networks, referred as $\ASYNC$, is discussed in our preliminary study~\cite{FL_v2x}, and illustrated in Fig. \ref{fig:v2v_fl_mle}.
Here, vehicles 
observe their queue length samples and
utilize the tail distribution of queue lengths at the vehicular transmitters over the whole edge network 
to optimize their transmission decisions 
such that the worst-case queue lengths are minimized while ensuring reliability in terms of queuing latency.
The analytical parametric model of the aforementioned tail distribution is obtained via \gls{evt}.
Naturally, the evaluation of above parameters is carried out by gathering all queue length samples at a central controller, the \gls{mec} server, with the additional costs of communication and computation overheads.
In contrast to the centralized approach, here, \gls{fl} is used to reduce the communication payload by allowing individual vehicles to learn the tail distribution by exchanging a simplified model (two gradient values) instead of their raw local queue length samples, i.e. enabling URLLC with the aid of ML at the edge devices.


The goal is thus to minimize the network-wide power consumption of a set of \glspl{vue} while ensuring low queuing latencies with high reliability. 
However, there still exists worst-case \glspl{vue} experiencing high latencies with a low probability whose performance losses are captured by extreme events pertaining to vehicles’ queue lengths exceeding a predefined threshold with non-negligible probability.
The principles of \gls{evt} characterize the tail distribution of the queue lengths exceeding a predefined threshold by a generalized Pareto distribution with two parameters scale and shape, respectively.
The concepts in \gls{mle} are used along \gls{fl} to estimate the scale and shape parameters of the queue tail distribution locally at each \glspl{vue} over the queue length samples.
Therein, occasionally, local estimations and the gradients of \gls{mle}, known as \emph{local model} at each \glspl{vue} are shared with the \gls{mec} server.
The \gls{mec} server does model averaging and shares the \emph{global model} with the \glspl{vue} to update their local estimations.
Using the knowledge of the tail distribution over the network, the transmit power of each \gls{vue} is optimized to reduce the worst-case queuing delays.

\begin{figure}[t!]
    \hspace{5pt}
	\subfigure[]{
		\includegraphics[width=.9\columnwidth]{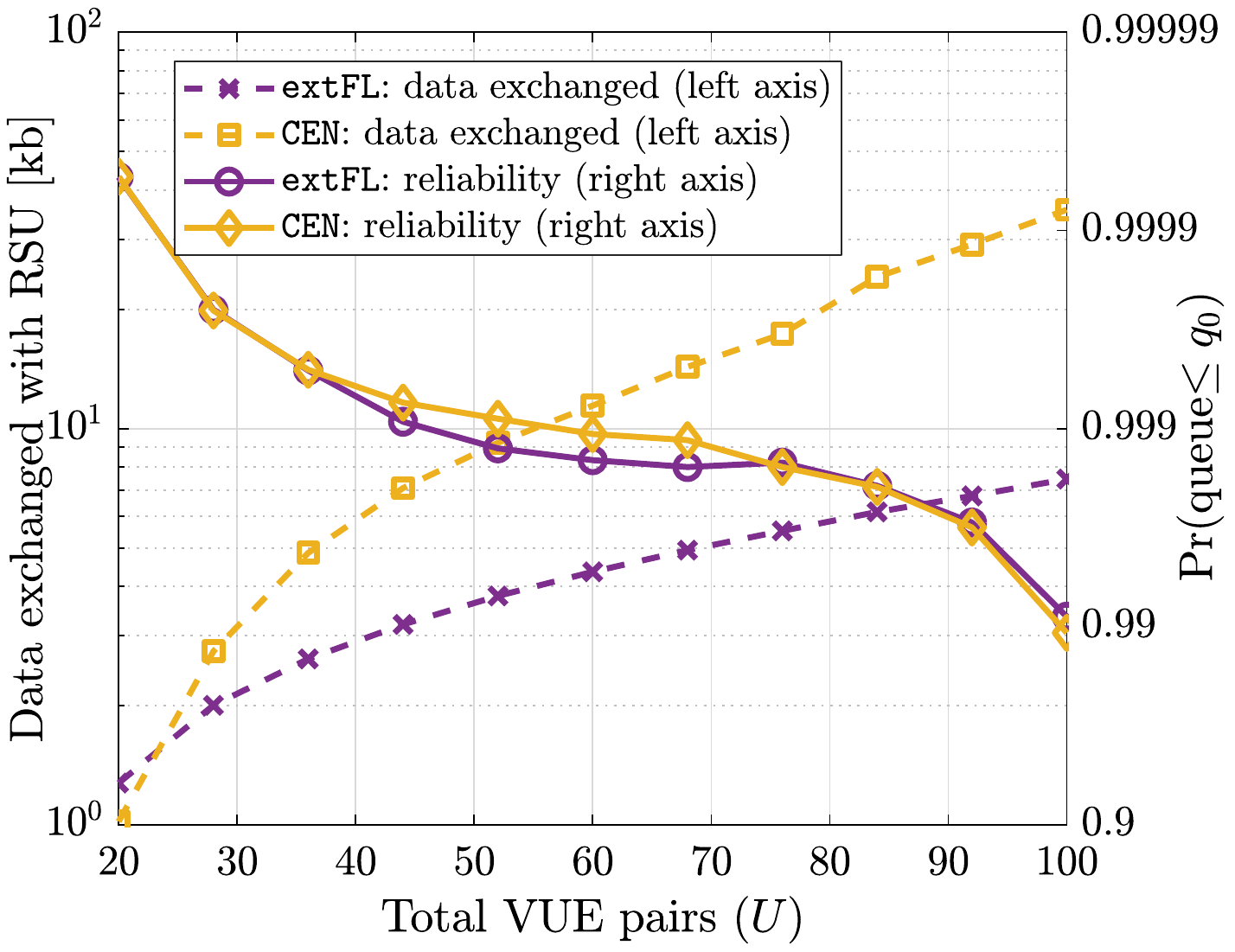}
		\label{fig:FLvsCEN}}
	\hspace*{18pt}
	\subfigure[]{
		\includegraphics[width=.8\columnwidth]{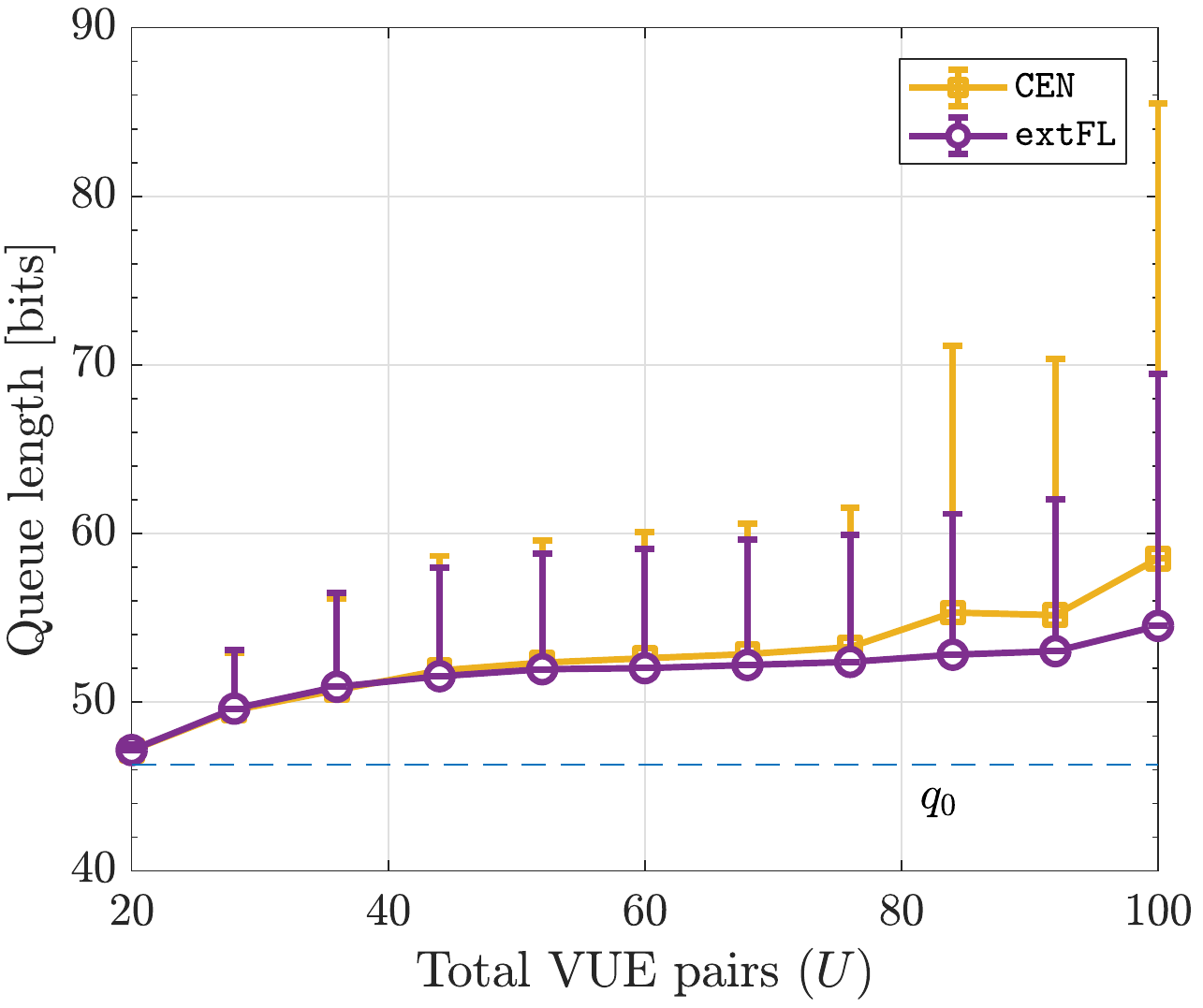}
		\label{fig:v2v_queues}}
	\caption{Comparison between $\CEN$ and $\ASYNC$. (a) The amount of data exchanged between RSU and VUEs (left axis) and the achieved reliability (right axis). (b) Mean and variance of the worst-case VUE queue lengths.}	
	\label{fig:accuracy}
\end{figure}

Fig. \ref{fig:FLvsCEN} compares the amount of data exchanged and the achieved V2V communication reliability of $\ASYNC$ with a centralized tail distribution estimation model, denoted as $\CEN$. 
%
%
Note that the $\CEN$ method requires all \glspl{vue} to upload all their queue length samples to the RSU and to receive the estimated GPD parameters.
In contrast, in \texttt{$\ASYNC$}, \glspl{vue} upload their locally estimated learning models and receive the global estimation of the model.
As a result, $\ASYNC$ yields equivalent or better end user reliability compared to $\CEN$ for denser networks while reducing the amount of data exchange among \glspl{vue}
and the RSU.
%
%
The worst-case \glspl{vue} queue lengths, i.e., queue lengths exceeding $\queueTH$, are compared in Fig. \ref{fig:v2v_queues}.
%
%
Here, the mean indicates the average queuing latency of the worst-case \glspl{vue} while the variance highlights the uncertainty of the latency.
As the number of \glspl{vue} increases, it can be noted that both the mean and variance in $\ASYNC$ are lower than the ones in $\CEN$.
The reason for above improvement is the reduced training latency in $\ASYNC$ over $\CEN$.

\begin{figure*}[t]
  \centering
  \includegraphics[width=0.9\textwidth]{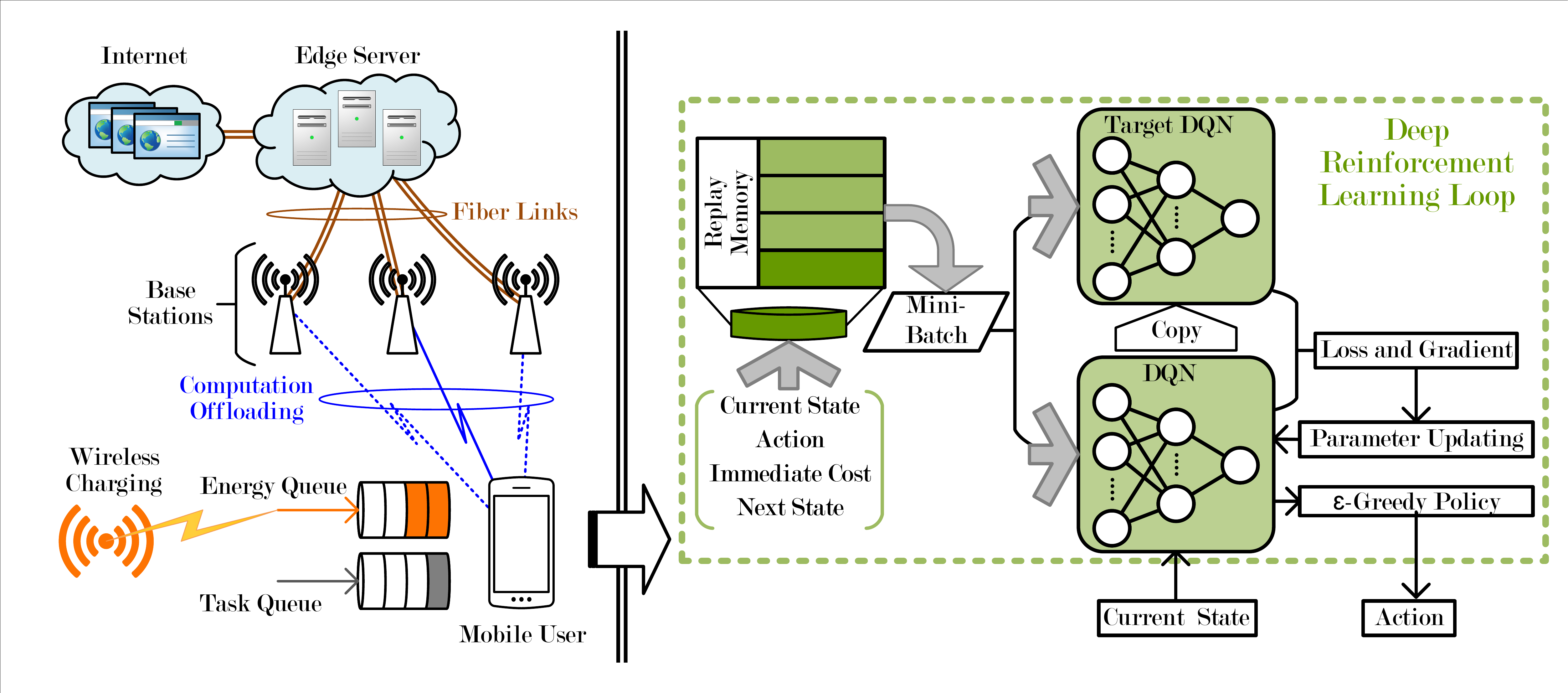}
  \caption{Illustration of deep reinforcement learning for mobile-edge computing performance optimization.}
  \label{fig:dqn}
\end{figure*}

\vspace*{0.2cm}
\label{usecaseDQN}
\usecase{Deep Reinforcement Learning for Optimized Edge Computing Task Offloading}:
The task offloading decision-making in edge computing networks is a challenging task in the presence of environmental dynamics. This situation is aggravated in ultra-dense networks, where solutions to break the \emph{curse of dimensionality} is desperately needed. 
In the works \cite{Chen1801,Chen1802}, a discrete-time Markov decision process was adopted to model the problem of expected long-term \gls{mec} performance optimization in an ultra-dense radio access network, where a number of \glspl{bs} are available for computation task offloading.
For a representative wireless charging enabled \gls{mue}, whether to execute an arriving computation task at the local mobile device or to offload the task for edge server execution via one of the \glspl{bs} should adapt to the environment dynamics in an intelligent manner.
These environment dynamics may consist of random computation task arrivals, time-varying communication qualities between the MU and the \glspl{bs} and the sporadic energy availability at the mobile device.
The challenges for the problem-solving lie in the lack of any a priori knowledge of any environment dynamic statistics along with the high dimensional state space.
A deep reinforcement learning technique shows the power of achieving an optimal solution.

More specifically, the objective of the \gls{mue} is to minimize an expected infinite-horizon discounted cost given by
\begin{align}\label{expeCost}
  Q(s, a) = \textsf{E}\!\!\left[ \sum_{t = 1}^\infty (\gamma)^{t - 1} \cdot c\!\left(s^t, a^t\right) | s^1 = s, a^1 = a\right],
\end{align}
where $\gamma \in [0, 1)$ is the discount factor, while the immediate cost $c\!\left(s^t, a^t\right)$ after performing an action $a^t$ under a state $s^t$ at each time slot $t$ takes into account the incurred task execution delay and the penalty of failing to process an arriving computation task.
Once we obtain the optimal $Q$-function, the optimal action $a^*$ can be made by the \gls{mue} following $a^* = \arg\min_a Q(s, a)$ under a state $s$.
Instead of using a conventional $Q$-learning to find the optimal $Q$-function, we resort to a \gls{dqn} \cite{Mnih15} $Q(s, a; \bm\theta)$ to approximate $Q(s, a)$ with $\bm\theta$ being the set of parameters of the neural network.
The procedure of the deep reinforcement learning for \gls{mec} performance optimization is briefly depicted as in Fig. \ref{fig:dqn}.

\begin{figure}[t]
  \centering
  \includegraphics[width=.95\columnwidth]{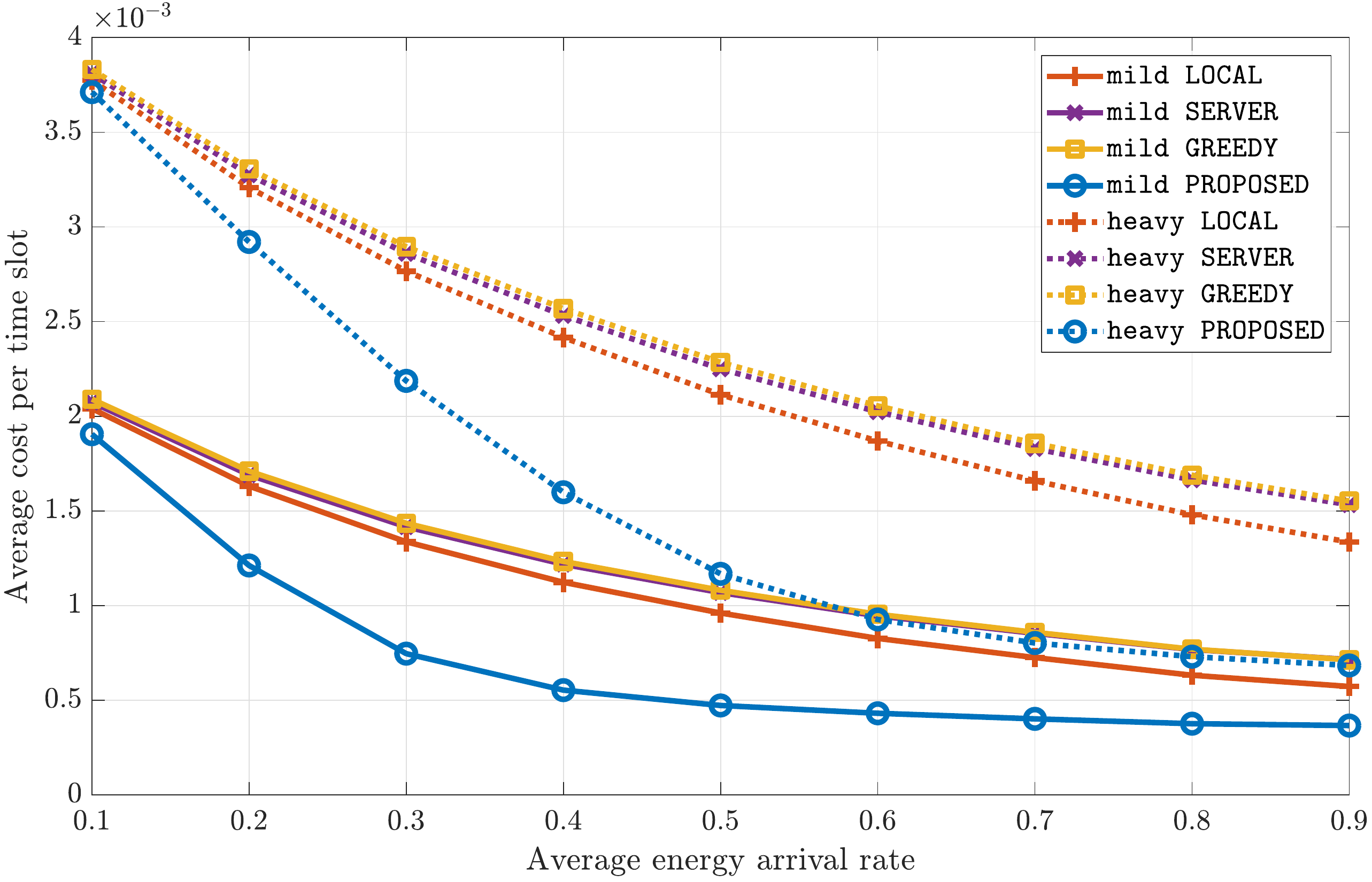}
  \caption{Average cost per time slot versus average energy arrival rate under \texttt{MILD} ($\rho\!=\!0.3$) and \texttt{HEAVY} ($\rho\!=\!0.5$) task arrival probabilities, respectively represented with solid and dashed lines.}
  \label{simu01}
\end{figure}

In Fig. \ref{simu01}, we compare the average cost performance from the \texttt{Proposed} deep reinforcement learning algorithm with three baselines: 1) \texttt{Local} -- Whenever a computation task arrives, the \gls{mue} executes it at the local mobile device using the queued energy units; 2) \texttt{Server} -- All arriving computation tasks are offloaded to the edge server for computing via the \glspl{bs} with the best communication qualities; and 3) \texttt{Greedy} -- When the computation task queue as well as the energy queue are not empty at a time slot, the \gls{mue} decides to execute the task locally or at the cloud to achieve the minimum immediate cost.
We configure a \gls{dqn} of one hidden layer with $512$ neurons.
The replay memory is assumed to have a capacity of $5000$ and we select the size of the mini-batch as $100$.
From Fig. \ref{simu01}, we can clearly see that compared to the baselines, the deep reinforcement learning algorithm realizes best performance in average cost.
A higher task arriving probability $\rho$ indicates a longer average task execution delay, hence a larger average cost.
As the average energy arrival rate increases, the average cost improves due to the decreased failure of processing an arriving computation task.

\begin{figure*}[ht!]
  \centering
  \includegraphics[width=\textwidth]{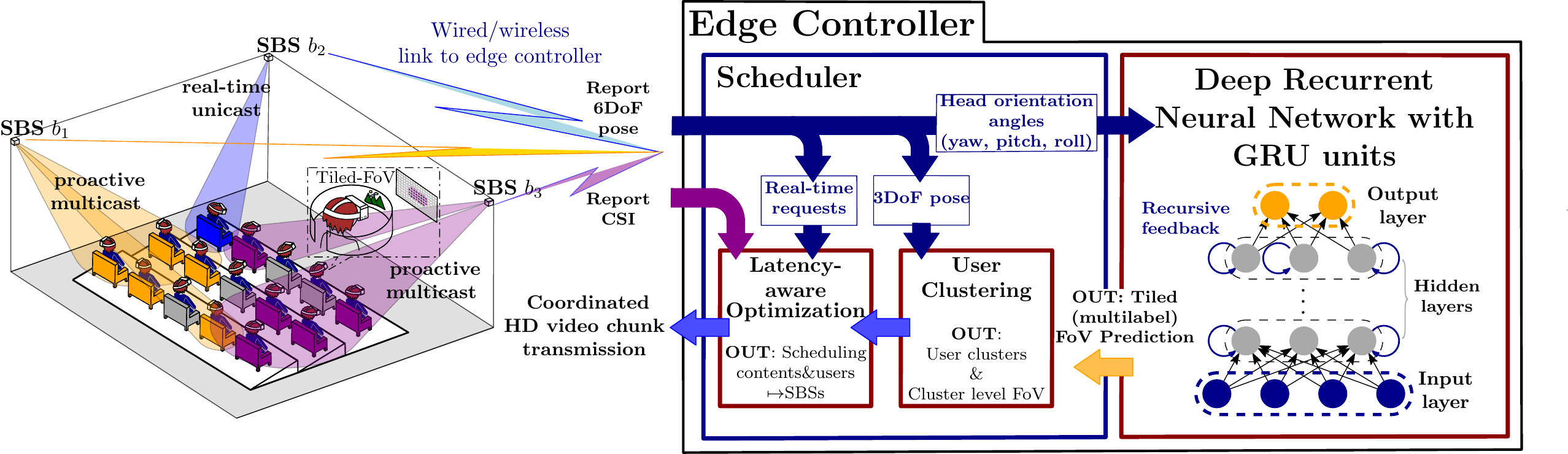}
  \caption{Operational structure and building blocks of the edge controller that coordinates the DRNN FoV prediction-aided proactive content quality adaptation for the \gls{mmwave} $360^{\circ}$ VR video streaming.}
  \label{fig:VR_theater_scenario}
\end{figure*}
\begin{figure*}[ht]
  \centering
  \includegraphics[width=\textwidth]{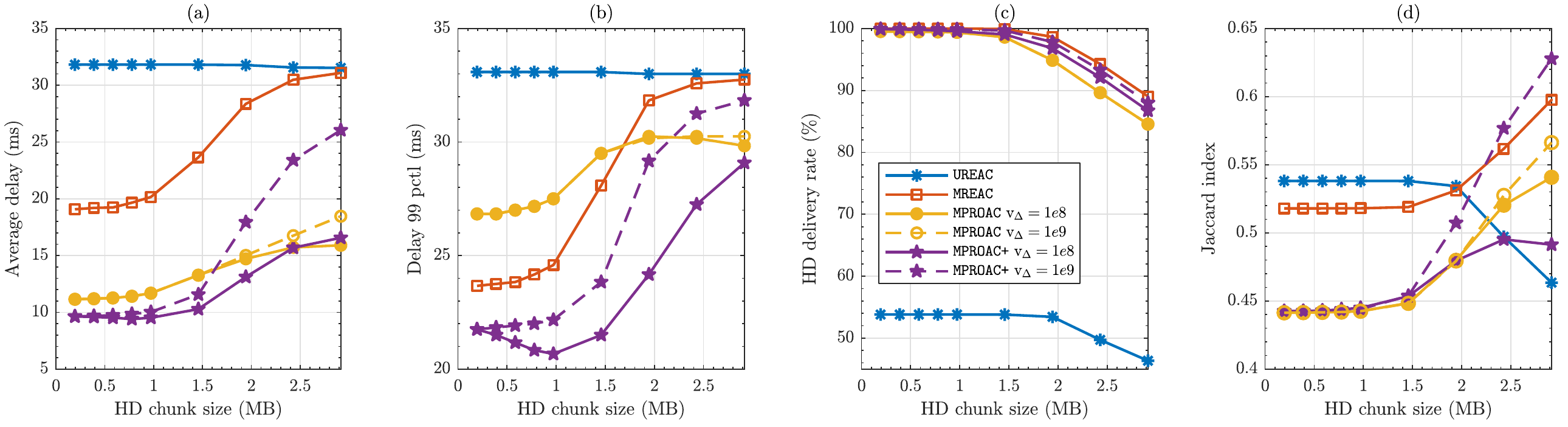}
  \caption{(a) Average delay, (b) $99$th percentile delay, (c) HD delivery rate and (d) Jaccard index performance in \texttt{sT-3v}, respectively, as a function of the HD chunk size, for $V\!=\!3$ videos, $K\!=\!2\times\!V$ clusters, $T_H\!=\!5$ frames, and Lyapunov trade-off $V_{\delta}\!=\!1\!\cdot\!10^8$ and $V_{\delta}\!=\!1\!\cdot\!10^9$.}
  \label{fig:360VR_chunkSize}
\end{figure*}
\vspace*{0.2cm}
\usecase{Edge ML Enabled 360$^\circ$ VR Multicast Transmission}\label{usecaseVR2}

Our previous work in~\cite{PerfectoVRStreaming} considered merging \gls{ml} and \gls{mmwave} multicasting to optimize the proactive wireless streaming of FoV-based \gls{hd} 360$^\circ$ videos in a multi-user \gls{vr} environment with low latency  guarantees. Hereof, the use of edge \gls{ml} to predict users' \gls{fov} in advance is pivotal to leverage inter-user correlations and curb the latency. These predicted correlations will ultimately drive both how contents are transmitted and the beamforming decisions at the \gls{mmwave} base stations. 

A VR theater scenario consisting of a network of VR users watching different HD $360^{\circ}$ VR videos 
streamed in the \gls{mmwave} band over a set of distributed \glspl{sbs} is studied. The \glspl{sbs} will report users' \gls{6dof} pose as well as \gls{csi} and produce multiple spatially orthogonal beams to serve shared \gls{fov} video content to groups of users (multicast) or a single beam (unicast) following the scheduling decisions adopted at the edge controller.
By optimizing video frame admission and user scheduling, the goal is to provide a highly reliable broadband service for \gls{vr} users that deliver \gls{hd} videos with a latency that is below the \gls{mtp} latency limits with very high probability. 

To achieve this proactive content transmission and perform a head movement pattern recognition predicting users' upcoming tiled-\gls{fov}, a sequential learning model based on \glspl{gru}~\cite{GRUseminal,GRUseminal2} is selected. Specifically,  \glspl{gru} are a form of \glspl{rnn} that include a double gating mechanism to govern the impact of past hidden states over the new output states and effectively tackle long-term dependencies. To that purpose, an architecture based on 2 layers of \gls{gru} cells with a hidden state size equal to 512 separated by a \gls{relu} activation are stacked. The output is then fed to a serial to parallel (S/P) layer and to a dense neural layer. Given the multi-label nature of the learning model, a sigmoid activation layer maps the $N$ sized dense output to the $N$ logits, one for each tile in the \gls{eqr} projection of the 360$^\circ$ \gls{vr} video frame, which are binarized with a cutoff layer such that 
\begin{equation}
\widehat {y}_{u,n}^{f_p}= 
\begin{cases}
1, & \sigma(\bm{W}_{d}\bm{h}_f^{(2)}+\bm{b}_d)_n\geq \gamma_{th},\\
0, & \text{otherwise},
\end{cases}
\end{equation}
where $\bm{W}_{d}$, $\bm{b}_d$ are the weights and biases of the dense fully-connected layer and $\gamma_{th}$ is the threshold value for the cutoff layer. The predicted FoV for a user $u$ and frame index \smash{$f_p=f+T_H$} is retrieved as $\widehat{\mathcal{N}}_{u}^{f_p}=\{ n\in [1,...,N]\hspace{-1mm}:\widehat {y}_{u,n}^{f_p}=1\}$.

Fig. \ref{fig:VR_theater_scenario} provides an overview of the building. The output of the DRNN is fed to a user clustering module and The former constitutes one of the inputs for a scheduler the Lyapunov Drift plus penalty approach.
In addition to our proposed scheme \texttt{MPROAC+}, the performance of three reference baselines with reactive unicast and multicast, and proactive multicast  transmission capabilities, correspondingly, \texttt{UREAC}, \texttt{MREAC}, and \texttt{MPROAC} is evaluated. Our proposed approach incorporates a penalty whereby quality is trade in exchange for not violating a maximum latency bound. 
For simulation purposes, a small size theatre with capacity for 50 users with  \glspl{sbs} are located at ceiling level in its upper 4 corners is selected. Fig. \ref{fig:360VR_chunkSize} evaluates the impact of the requested \gls{hd} video quality by representing the average and $99^{th}$ percentile delay, the HD delivery rate and Jaccard index measured while 30 users watch one out of the 3 available VR videos for an increasing requested video chunk size. 

Fig. \ref{fig:360VR_chunkSize} clearly shows the tradeoff between frame delay and \gls{hd} streaming rate. As the chunk size increases, the average and 99th percentile delays increase for the different schemes. Moreover, comparing \texttt{UREAC} with the other schemes, it is shown that multicasting brings $40-50\%$ increase in the \gls{hd} rate and $33-70\%$ latency reduction through the utilization of shared \glspl{fov} of different users. By delivering the predicted frames in advance, both the  \texttt{MPROAC} and \texttt{MPROAC+} minimize the average delay without sacrificing the HD quality rate. Moreover, our proposed \texttt{MPROAC+} scheme is shown to also keep the worst delay values bounded due to imposing the constraint over the latency.

The tradeoff between frame delay and quality is further illustrated the results for different values of the Lyapunov parameter $V_\delta$ are compared; as $V_\delta$ increases, the scheduling algorithm prioritizes maximizing users' \gls{hd} delivery rate, whereas at lower values of the scheduler prioritizes keeping the delay bounded with high probability. This comes at the expense of having lower \gls{hd} delivery rate.

Lastly, the Jaccard similarity in Fig.~\ref{fig:360VR_chunkSize}(d) illustrates the tradeoffs between effective vs. transmitted contents. At low traffic loads, the Jaccard index is low, which is due to the large amount of excess data delivered due to transmitting an estimated user/cluster level \gls{fov}. As the traffic load increases, the proactive schemes transmit more real-time frames, which increases the Jaccard index. The Jaccard index decreases again at higher traffic loads as the effect of missed frames increases (once the average delay is close to reaching the deadline, as can be seen in Fig.~\ref{fig:360VR_chunkSize}(a)).

\begin{figure*}[ht]
  \centering
  \includegraphics[width=0.70\textwidth]{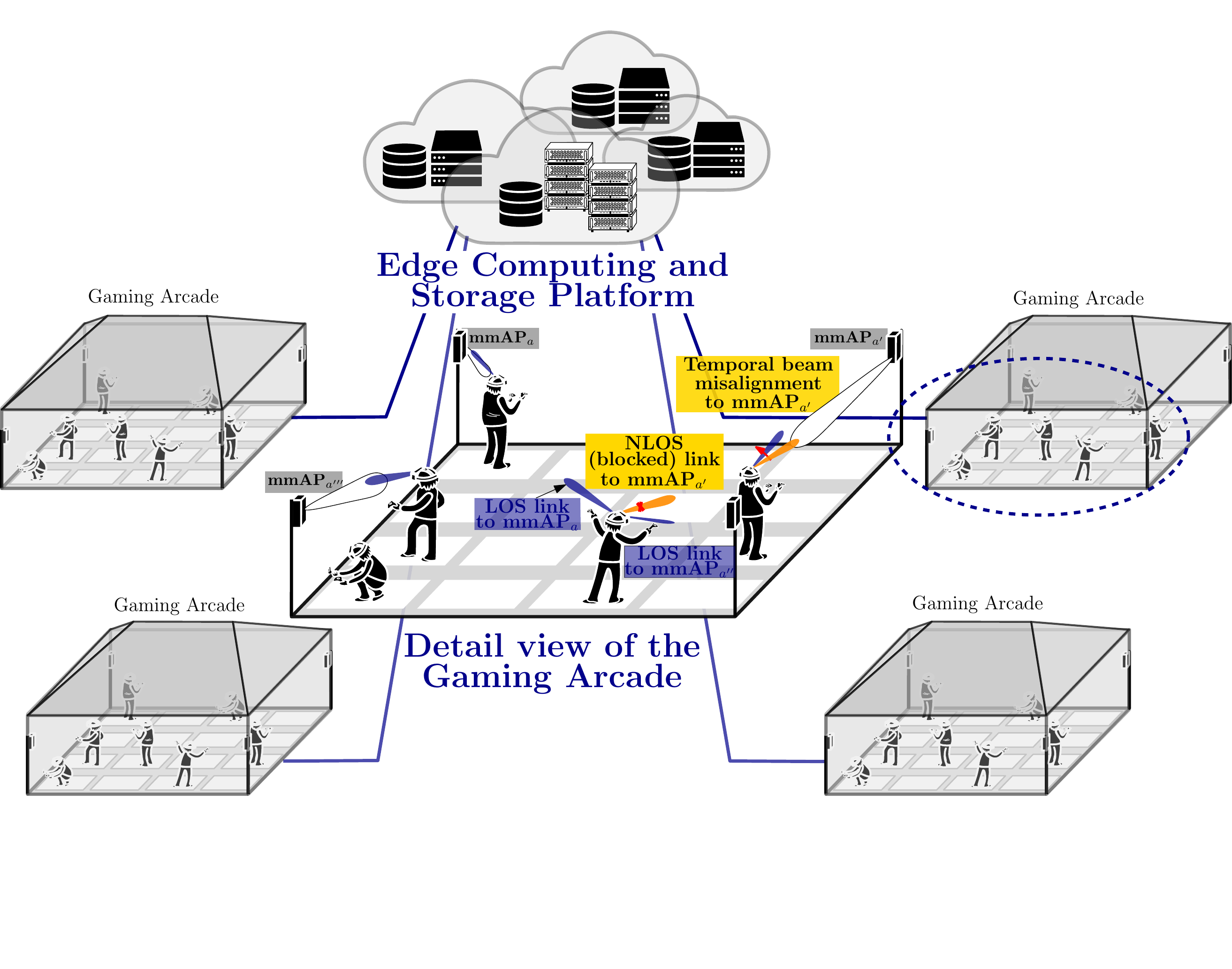}
  \caption{Representation of a group of VR gaming arcades where HD frame computation is offloaded to a MEC platform such that input actions of the \glspl{vrp} might impact the virtual environment shown to a subset of the remaining \glspl{vrp}. The detailed view of the bottom arcade also illustrates several \gls{los} and \gls{nlos} \gls{mmwave} link states e.g., link blockage and \gls{vrp} and \gls{mmap} beam misalignment.}
  \label{fig:VR_scenario}
\end{figure*}

\vspace*{0.2cm}
\usecase{MEC Enabled Multi-User VR Gaming Arcade}\label{usecaseVR1}
We consider a practical use case of wireless \gls{vr} to deliver a low latency service to multi-user scenario of users playing \gls{vr} video games in a gaming arcade.
This scenario, that is fully detailed in our previous work~\cite{conf:Elbamby2018-WCNC}, is highly demanding due to the tight latency tolerance in \gls{vr} as well as the state dynamics of the user due to the game-specific actions taken by themselves or by other players that affect what content should be shown to them.
The users are served wirelessly through multiple \glspl{mmap} wired to edge computing and storage servers.
These servers receive the users’ \gls{3d} location coordinates, their \gls{3d} pose that consists of roll, pitch, and yaw angles, and their game-related actions. 
The servers will render the corresponding frames in \gls{hd} resolution and deliver it wirelessly to users.
Hence, the latency consists of the processing latency at the server and the communication latency to deliver the \gls{hd} frames expressed as

\begin{equation}
D_{uf}(t)=\xi_{fu}(D_{uf}^{\textrm{cp}}(t)+D_{uf}^{\textrm{cm}}(t)+\text{\ensuremath{\tau}}_{\textrm{EP}}),
\end{equation}
where $\xi_{fu}$ represents a binary indicator that equals $1$ when the
\gls{hd} video frame is delivered to \gls{vrp} $u$ and equals $0$ if the \gls{lq}
frame is delivered, $D_{uf}^{\textrm{cp}}$ and $D_{uf}^{\textrm{cm}}$
are the computing and communication delays of HD frame $f$ initiated
from user $u$, and $\text{\ensuremath{\tau}}_{\textrm{EP}}$ is the
processing latency which accounts for the edge server processing,
storage processing, and the UL transmission of user pose and action
data. 
Let the computing delay $D_{uf}^{\textrm{cp}}$ be expressed as follows: \vspace{-0.1cm}

\begin{equation}
D_{uf}^{\textrm{cp}}(t)=\biggl(\frac{\kappa L_{fu}^{\textrm{HD}}}{c_{e}}+W_{uf}(t)\biggr)z_{fu}(t)(1-y_{fu}(t)),
\end{equation}
where $c_{e}$ is the computation capability of edge server $e$, $z_{fu}(t)$ and $y_{fu}(t)$ indicate that the video frame $f$ of user $u$ is scheduled for computing, and is cached in the fog network
at time instant $t$, respectively, and $W_{uf}$ is the computation waiting time of HD frame $f$ of user $u$ in the service queue, defined as $Q(t)$. Furthermore, let the communications delay $D_{uf}^{\textrm{cm}}$ be given as
\vspace{-0.4cm}

\begin{equation}
D_{uf}^{\textrm{cm}}(t)\hspace{-0.7mm}=\hspace{-0.7mm}\arg\min_{d_{u}}\hspace{-2mm}\sum_{t'=D_{uf}^{\textrm{cp}}(t)+1}^{D_{uf}^{\textrm{cp}}(t)+d_{u}}\hspace{-0.5mm}\biggl(T_{t}r_{u}(t')\geq L_{fu}^{\textrm{HD}}\biggr),
\end{equation}
where the \emph{$\arg\min$} function is to find the minimum number
of time slots needed for the video frame $f$ to be delivered.

Here, we study two enablers to minimize the latency and boost the reliability of the \gls{vr} gaming experience. For the computing latency, we investigate how prior knowledge of users' future pose using prediction methods affects the computing the latency. We leverage results from previous works as in~\cite{Qian16} that state that the user's future pose in the next hundreds of milliseconds can be predicted with high accuracy to proactively predict, render, and cache the users’ upcoming frames, subject to computation and storage resource availability.
For the communication parts, the use of \gls{mxconn} is considered to associate a user with more than one \glspl{mmap} if the \gls{sinr} with its serving \gls{mmap} falls below a given threshold. Specifically, \gls{sfn} operation is considered where multiple \gls{mmap} use the same frequency and time resource to transmit to the intended user.

\begin{figure}[t]
  \centering
  \includegraphics[width=\columnwidth]{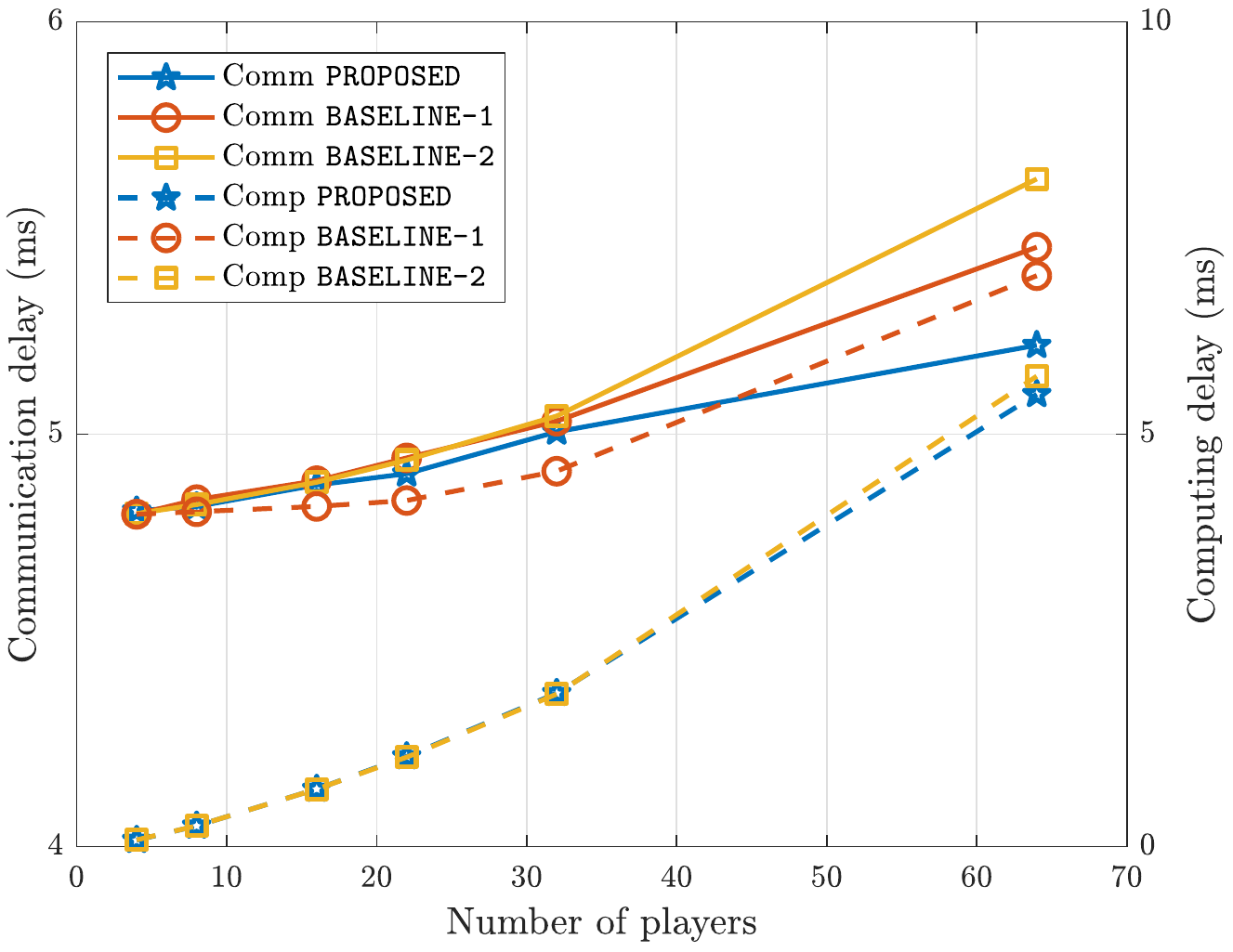}
  \caption{The communication delay (solid lines) and  computing delay (dashed lines) for different schemes as the number of players varies for an arcade of 16 mmAPs, each equipped with an edge computing unit.}
  \label{fig:VR}
\end{figure}

Fig. \ref{fig:VR} compares the communications and computing latency of our $\Proposed$ scheme that considers both enablers of proactive computing and \gls{mxconn}, with $\FirstBaseline$ that does not have either of the two enablers, and $\SecondBaseline$ that considers only proactive computing. By looking into the computing latency in Fig. \ref{fig:VR}, we can see that the schemes with proactive computing significantly minimizes the computing latency, whereas a look at the communication latency shows the gain achieved using \gls{mxconn}. Comparing the communication latency of $\FirstBaseline$ and $\SecondBaseline$ also shows that the proactive computing, that improved the computing performance, also slightly increases the communication latency. This is due to having to send additional data due to the errors in prediction, in which the correct data has to be retransmitted in real time.  

\section{Conclusions and Future Outlook} 
 
Edge computing is an essential component of future wireless networks, in which several challenges need to be overcome to realize the vision of ultra-reliable and low-latency edge computing.  
Chief to this vision is leveraging multiple high reliability and low-latency enablers applied for different types of services and use cases. In this article, we have discussed edge networking services and examined key enablers to achieve low-latency and high reliability networking. Moreover, we showcased how the network resources can be optimized for a selection of use cases characterized by their shared need for edge networking.
As the vision of \gls{5g} starts to materialize beyond its initial inception towards imminent first commercial deployments, we envision a realization of edge computing hand in hand with the development of \gls{urllc} and distributed \gls{ai} able to deal with dynamic and heterogeneous environments, provide seamless computing, content, and control services, while preserving data privacy and security.


\bibliographystyle{IEEEtran}
\bibliography{IEEEabrv,mec_bib}
\vspace*{-0,5cm}
\begin{IEEEbiography}[{\includegraphics[width=1in,height=1.25in,clip,keepaspectratio]{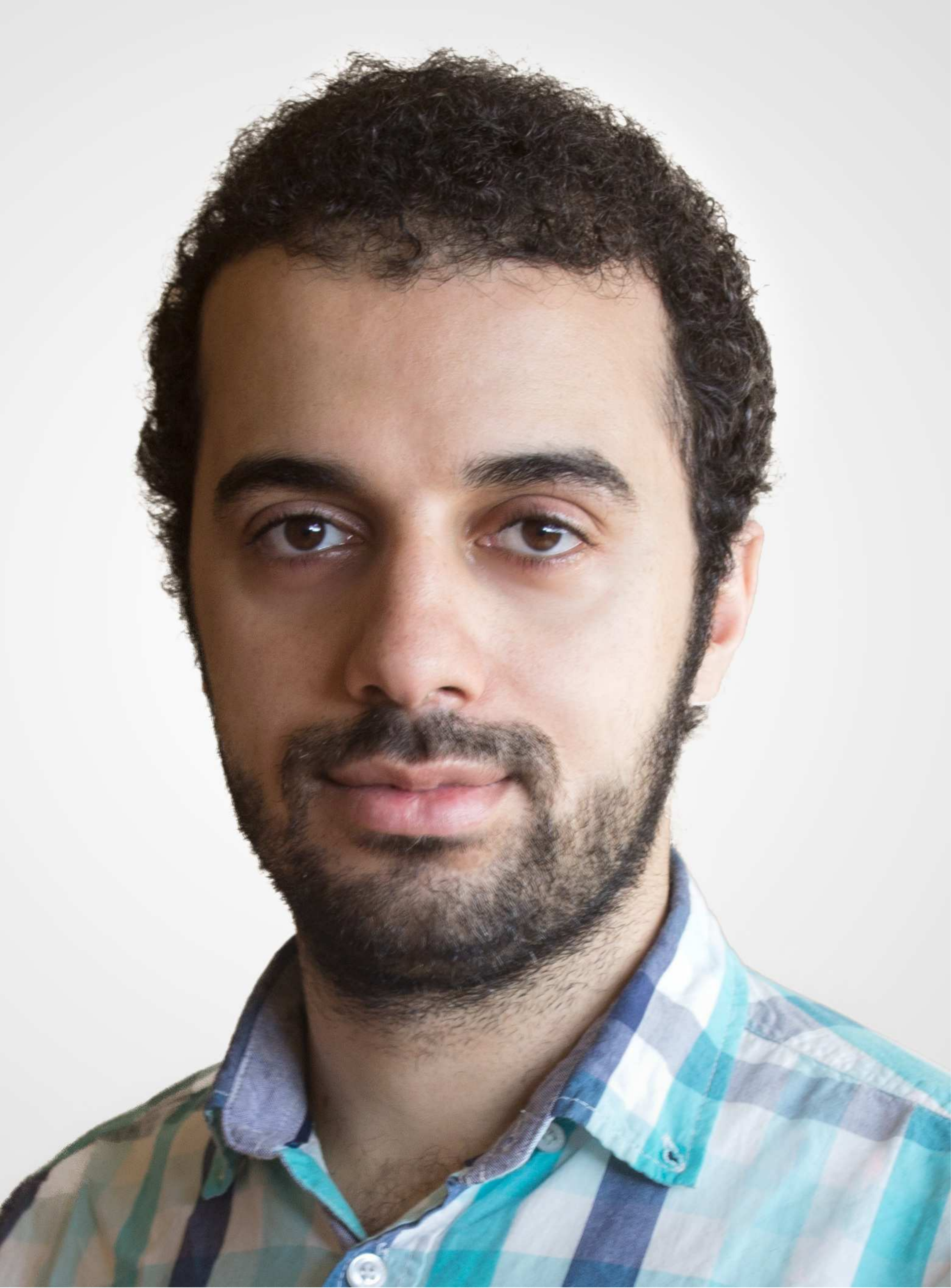}}]{Mohammed S. Elbamby} received the B.Sc. degree (Hons.) in Electronics and Communications Engineering from the Institute of Aviation Engineering and Technology, Egypt, in 2010, and the M.Sc. degree in Communications Engineering from Cairo University, Egypt, in 2013. He is currently pursuing the Dr.Tech. degree with the University of Oulu. After receiving the M.Sc. degree, he joined the Centre for Wireless Communications, University of Oulu. His research interests include resource optimization, uplink and downlink configuration, fog networking, and caching in wireless cellular networks. He received the Best Student Paper Award from the European Conference on Networks and Communications in 2017.
\end{IEEEbiography}
\vspace*{-0,5cm}
\begin{IEEEbiography}[{\includegraphics[width=1in,height=1.25in,clip,keepaspectratio]{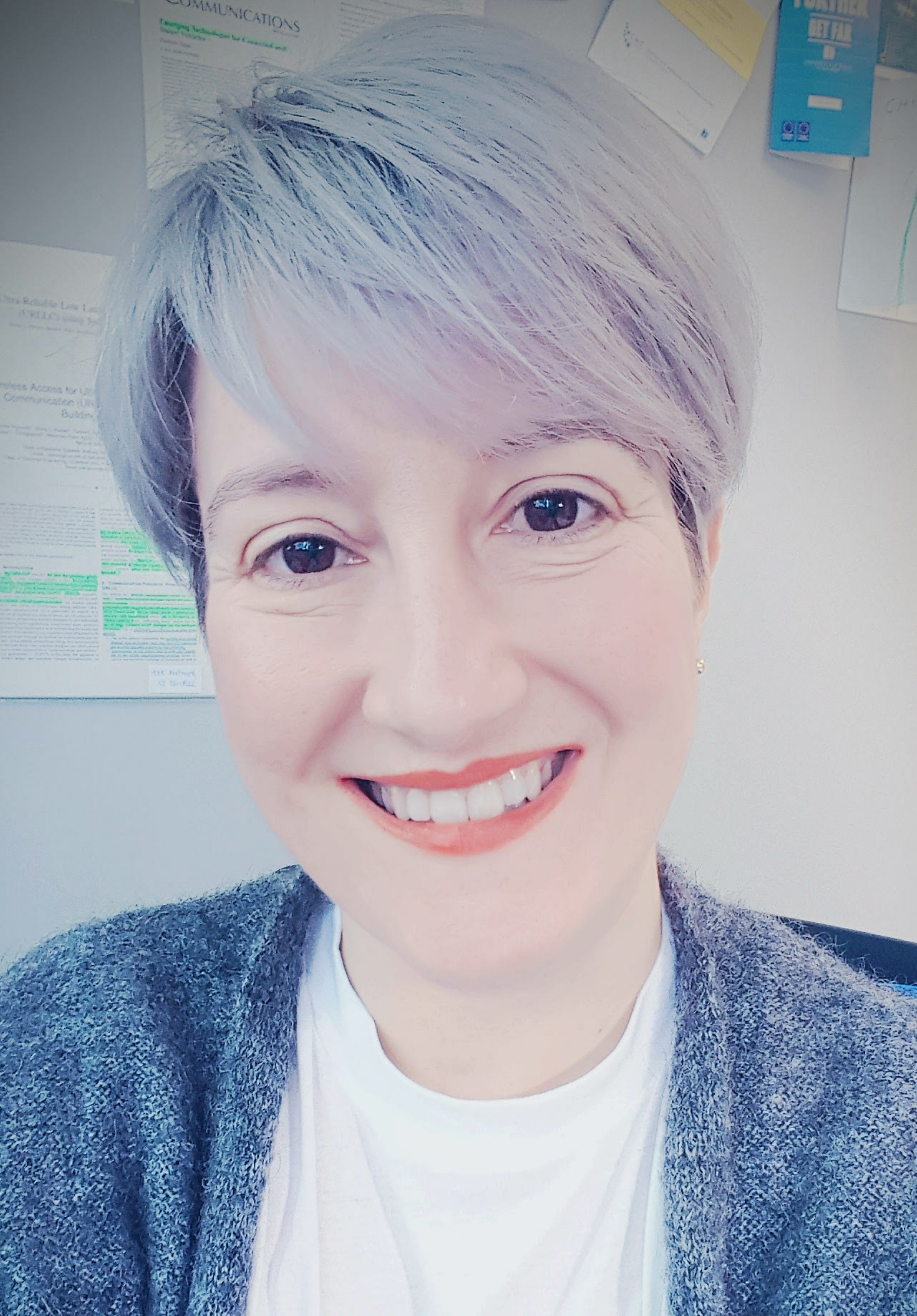}}]{Cristina Perfecto} (S'15) received her B.Sc. and M.Sc. in Telecommunication Engineering from the University of the Basque Country (UPV/EHU) in 2000. She is currently a college associate professor with the Department of Communications Engineering at the UPV/EHU. Her research interests lie on millimeter wave communications and in the application of machine learning in 5G networks. She is currently working towards her Ph.D. focused on the application of multidisciplinary computational intelligence techniques in radio resource management for 5G.
\end{IEEEbiography}
\vspace*{-0,5cm}
\begin{IEEEbiography}[{\includegraphics[width=1in,height=1.25in,clip,keepaspectratio]{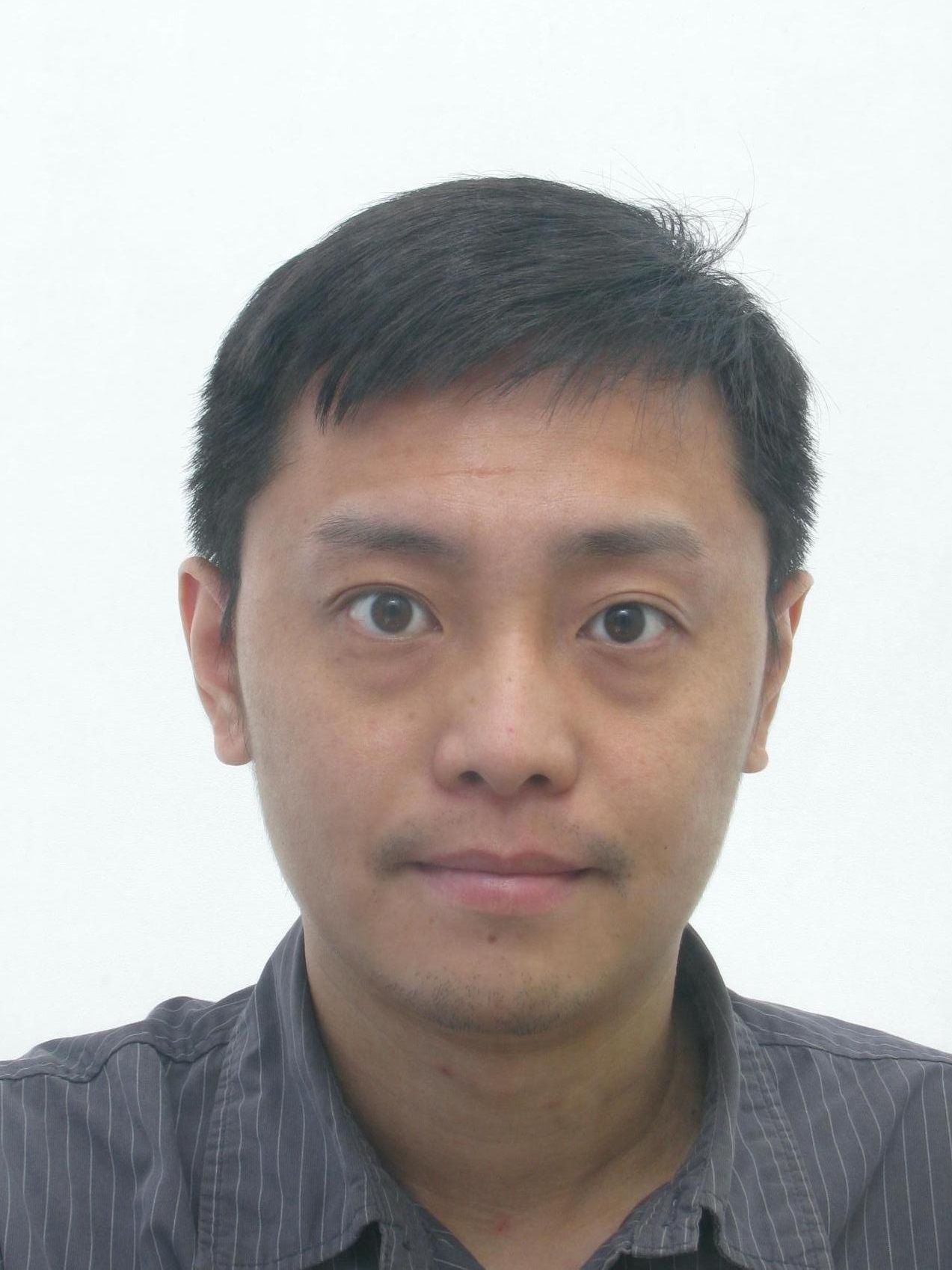}}]{Chen-Feng Liu} (S'17) received the B.S. degree from National Tsing Hua University, Hsinchu, Taiwan, in 2009, and the M.S. degree in communications engineering from National Chiao Tung University, Hsinchu, in 2011. He is currently pursuing the Ph.D. degree with the University of Oulu, Oulu, Finland. In 2012, he joined Academia Sinica as a Research Assistant. In 2014, he was a Visiting Researcher with the Singapore University of Technology and Design, Singapore. He was also a Visiting Ph.D. Student with the University of Houston, Houston, TX, USA, and New York University, New York, NY, USA, in 2016 and 2018, respectively. His current research interests include 5G communications, mobile edge computing, ultra-reliable low latency communications, and wireless artificial intelligence.
\end{IEEEbiography}
\vspace*{-0,5cm}
\begin{IEEEbiography}[{\includegraphics[width=1in,height=1.25in,clip,keepaspectratio]{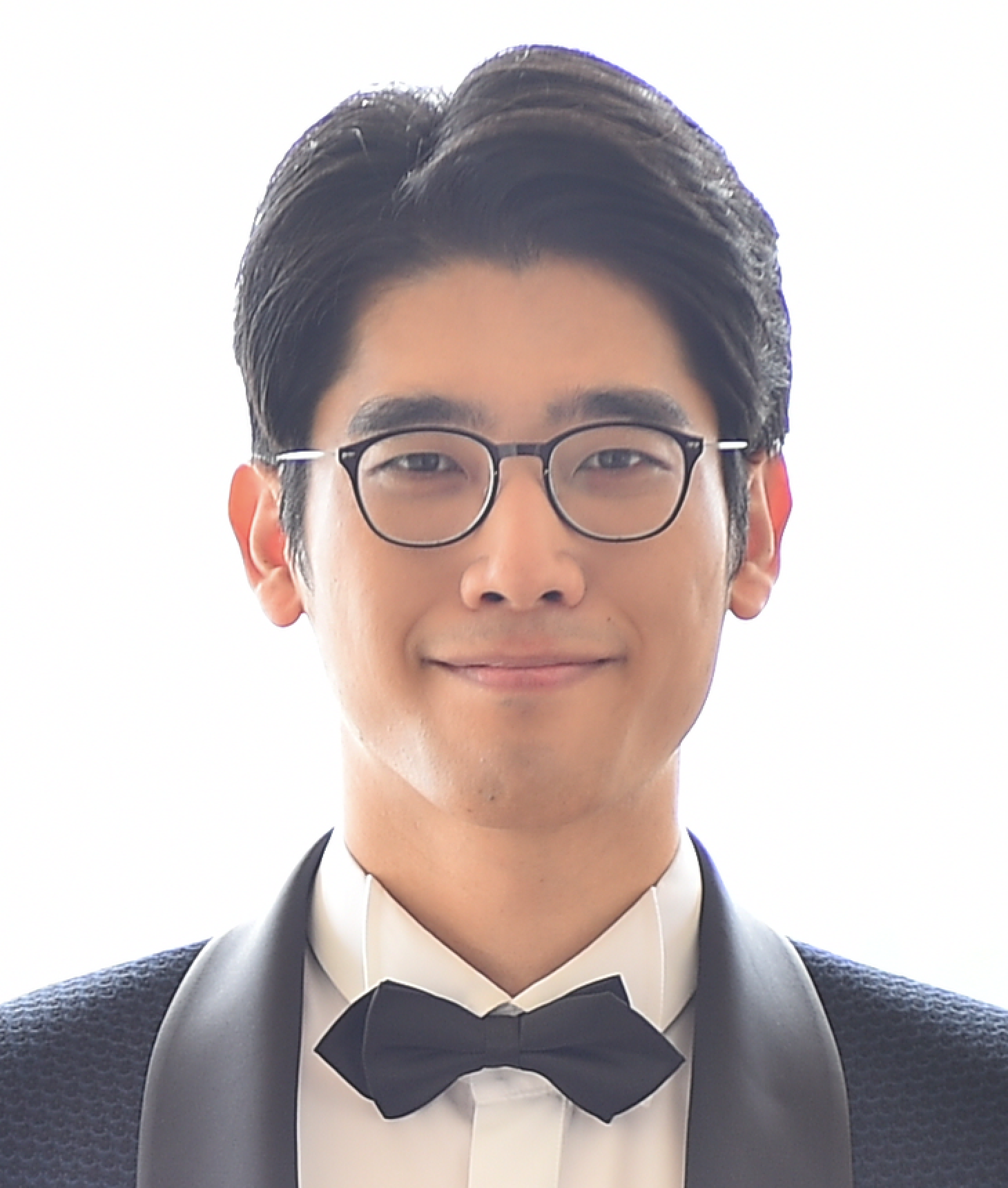}}]{Jihong Park} received the B.S. and Ph.D. degrees from Yonsei University, Seoul, South Korea, in 2009 and 2016, respectively. From 2016 to 2017, he was a Post-Doctoral Researcher with Aalborg University, Denmark. He was a Visiting Researcher with Hong Kong Polytechnic University; KTH, Sweden; Aalborg University, Denmark; and New Jersey Institute of Technology, USA, in 2013, 2015, 2016, and 2017, respectively. He is currently a Post-Doctoral Researcher with the University of Oulu, Finland. His research interests include ultra-dense/ultra-reliable/massive-MIMO system designs using stochastic geometry and network economics. His papers on tractable ultra-dense network analysis received the IEEE GLOBECOM Student Travel Grant in 2014, the IEEE Seoul Section Student Paper Contest Bronze Prize in 2014, and the 6th IDIS-ETNEWS (The Electronic Times) Paper Contest Award sponsored by the Ministry of Science, ICT, and Future Planning of Korea.
\end{IEEEbiography}
\vspace*{-0,5cm}
\begin{IEEEbiography}[{\includegraphics[width=1in,height=1.25in,clip,keepaspectratio]{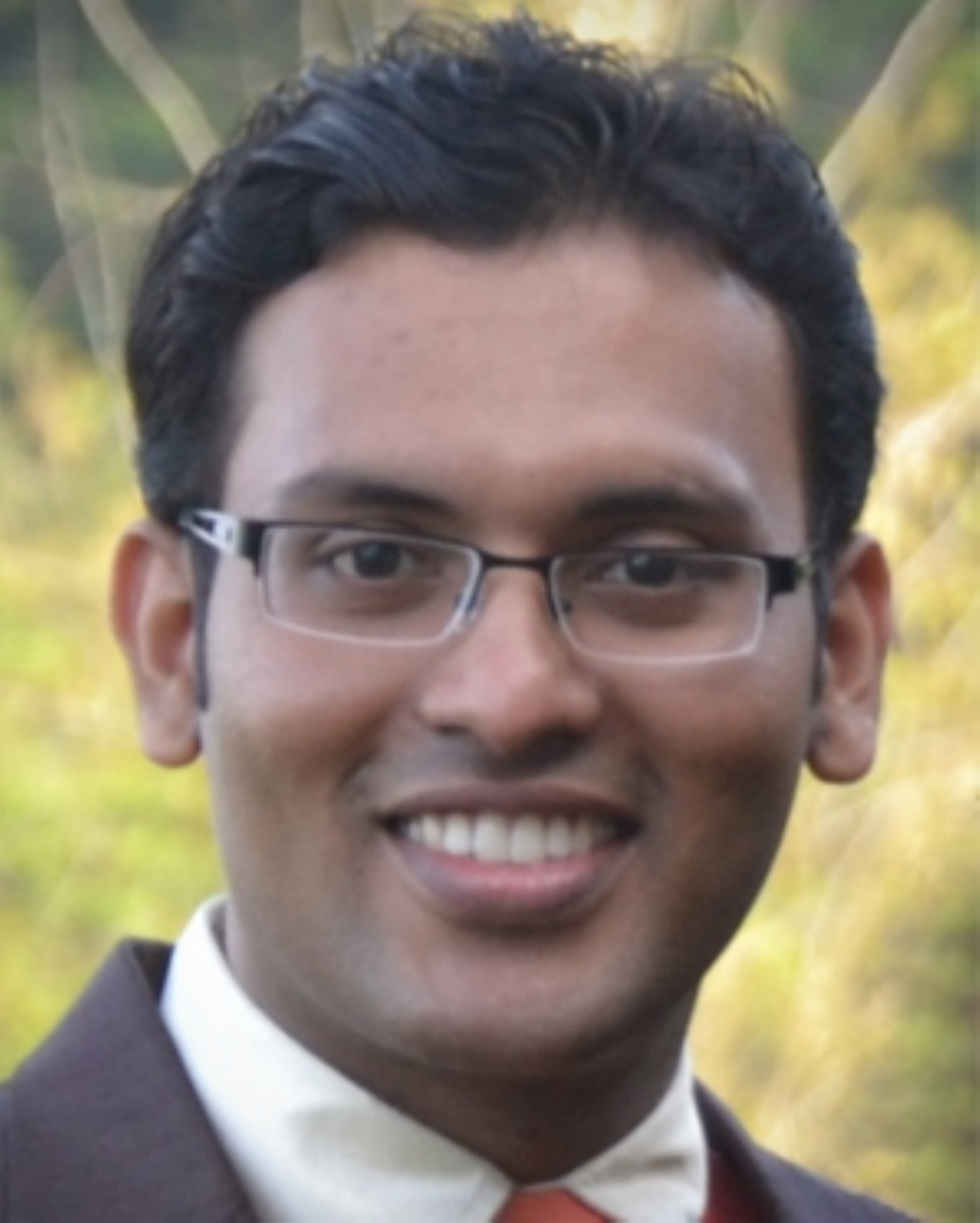}}]{Sumudu Samarakoon} (S'08-AM'18) received his B. Sc. Degree (Hons.) in Electronic and Telecommunication Engineering from the University of Moratuwa, Sri Lanka in 2009, the M. Eng. degree from the Asian Institute of Technology, Thailand in 2011, and Ph. D. degree in Communication Engineering from University of Oulu, Finland in 2017. He is currently working in Centre for Wireless Communications, University of Oulu, Finland as a post doctoral researcher. His main research interests are in heterogeneous networks, small cells, radio resource management, reinforcement learning, and game theory. In 2016, he received the Best Paper Award at the European Wireless Conference and Excellence Awards for innovators and the outstanding doctoral student in the Radio Technology Unit, CWC, University of Oulu.
\end{IEEEbiography}
\balance
\begin{IEEEbiography}[{\includegraphics[width=1in,height=1.25in,clip,keepaspectratio]{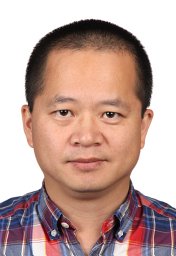}}]{Xianfu Chen} received his Ph.D. degree in Signal and Information Processing, from the Department of Information Science and Electronic Engineering at Zhejiang University, Hangzhou, China, in March 2012. He is currently a Senior Scientist with the VTT Technical Research Centre of Finland Ltd, Oulu, Finland. His research interests cover various aspects of wireless communications and networking, with emphasis on network virtualization, software-defined radio access networks, green communications, centralized and decentralized resource allocation, dynamic spectrum access, and the application of machine learning to wireless communications.
\end{IEEEbiography}
\begin{IEEEbiography}[{\includegraphics[width=1in,clip,height=1.25in,keepaspectratio]{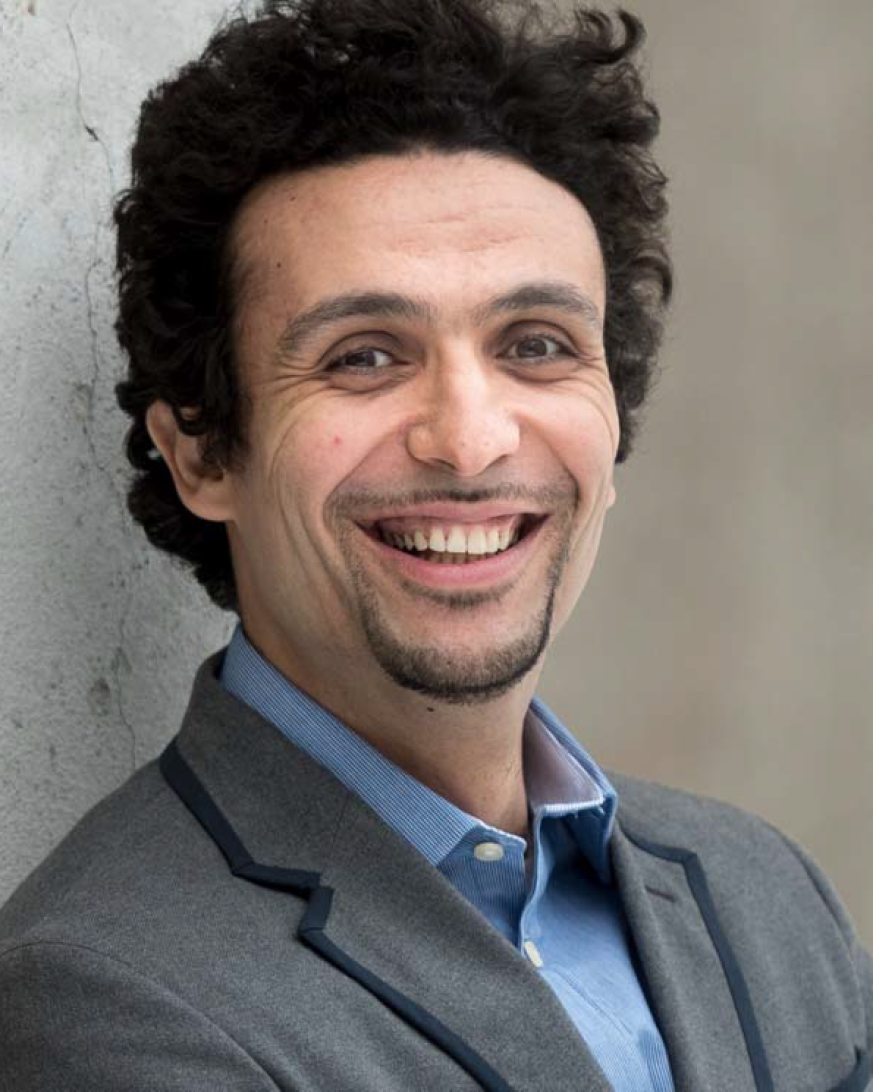}}]{Mehdi Bennis} (S'07-AM'08-SM'15)  received his M.Sc. degree in Electrical Engineering jointly from the EPFL, Switzerland and the Eurecom Institute, France in 2002. From 2002 to 2004, he worked as a research engineer at IMRA-EUROPE investigating adaptive equalization algorithms for mobile digital TV. In 2004, he joined the Centre for Wireless Communications (CWC) at the University of Oulu, Finland as a research scientist. In 2008, he was a visiting researcher at the Alcatel-Lucent chair on flexible radio, SUPELEC. He obtained his Ph.D. in December 2009 on spectrum sharing for future mobile cellular systems. Currently Dr. Bennis is an Associate Professor at the University of Oulu and Academy of Finland research fellow. His main research interests are in radio resource management, heterogeneous networks, game theory and machine learning in 5G networks and beyond. He has co-authored one book and published more than 100 research papers in international conferences, journals and book chapters. He was the recipient of the prestigious 2015 Fred W. Ellersick Prize from the IEEE Communications Society, the 2016 Best Tutorial Prize from the IEEE Communications Society and the 2017 EURASIP Best paper Award for the Journal of Wireless Communications and Networks. Dr. Bennis serves as an editor for the IEEE Transactions on Wireless Communication.
\end{IEEEbiography}
\vfill
\end{document}